\documentclass[review]{elsarticle}

\usepackage{lineno}
\usepackage{hyperref}

\usepackage{amssymb,amsbsy,amsmath,amsthm,amsfonts,amscd}
\usepackage{relsize}
\usepackage{xcolor}
\usepackage{graphicx}

\usepackage{subfig}
\usepackage{float}
\usepackage{subcaption}
\usepackage{multirow,tabularx}

\usepackage{enumitem}

\usepackage{algorithm2e}
\usepackage{algcompatible}

\usepackage{xcolor}

\newcommand{\vectb}{\boldsymbol}

\journal{Journal of Non-Newtonian Fluid Mechanics}
\biboptions{sort&compress}

\begin{document}

\begin{frontmatter}

\title{An insight into parameter identifiability issues in the Carreau-Yasuda model: A more consistent rheological formulation for shear-thinning non-Newtonian inelastic fluids}

\author[indirizzoTVdicii,indirizzoSAP]{Gianluca Santesarti\corref{mycorrespondingauthor}}
\cortext[mycorrespondingauthor]{Corresponding author}
\ead{santesarti@ing.uniroma2.it}

\author[indirizzoTVdicii]{Michele Marino}
\ead{m.marino@ing.uniroma2.it}

\author[indirizzoGSSI]{Francesco Viola}
\ead{francesco.viola@gssi.it}

\author[indirizzoGSSI,indirizzoTVind,indirizzoNL]{Roberto~Verzicco}
\ead{verzicco@uniroma2.it}

\author[indirizzoTVdicii,indirizzoBR]{Giuseppe Vairo}
\ead{vairo@ing.uniroma2.it}

\address[indirizzoTVdicii]{University of Rome Tor Vergata, Department of Civil Engineering and Computer Science Engineering, 00133 Rome, Italy}
\address[indirizzoSAP]{Sapienza University of Rome, Department of Mechanical and Aerospace Engineering}
\address[indirizzoGSSI]{Gran Sasso Science Institute, L'Aquila, 67100, Italy}
\address[indirizzoTVind]{University of Rome Tor Vergata, Department of Industrial Engineering, 00133 Rome, Italy}
\address[indirizzoNL]{Physics of Fluids Group, Max Planck Center for Complex Fluid Dynamics, MESA+ Institute and J. M. Burgers Centre for Fluid Dynamics, University of Twente, P.O. Box 217, 7500AE Enschede, Netherlands}
\address[indirizzoBR]{Universidade de Brasília, Department of Mechanical Engineering, 70910-900 Brasília, DF, Brazil}

\begin{abstract}
The Carreau-Yasuda rheological model is widely employed in both research and
industrial applications to describe the shear-thinning behaviour of non-Newtonian
inelastic fluids.
However, the model parameter traditionally employed to characterize the
shear thinning response exhibits only
a weak correlation with the actual shear thinning rate observed in experimental
data. This limitation leads to intrinsic identifiability issues, which may result in
misleading physical interpretations of the model parameters and unreliable flow
predictions.
Aiming to contribute to overcoming these issues, this paper introduces a novel heuristic rheological formulation for shear-thinning
non-Newtonian inelastic fluids, as an alternative to the Carreau-Yasuda model.
Analytical results and exemplary numerical case
studies demonstrate that the proposed formulation is based on physically meaningful
model parameters, whose identifiability is not compromised by the key limitations
of the Carreau-Yasuda model. The new approach allows for effective
parameter estimation through a straightforward direct identification strategy,
eliminating the need for inverse identification procedures based on nonlinear
regression techniques.
Moreover, the proposed formulation naturally enables the
identication of two Carreau numbers based on the two characteristic shear rates
of the fluid.
\end{abstract}
\end{frontmatter}

\nolinenumbers
\section{Introduction}
Fluids used in a wide range of advanced applications exhibit a complex non-Newtonian
and nonlinear rheological behaviour, characterized by a significant dependence
of viscosity on shear rate and/or shear rate history. Typical examples can be
found in biomedical engineering \cite{johnston2004non, amorim2021insights, sauty2022enabling},
tribology and industrial processes for the production of lubricants and paints \cite{chhabra2011non},
plastic polymer manufacturing \cite{bird1987dynamics},  food engineering \cite{rao2010rheology}.
Among the different types of these complex fluids, in the following reference
is made to the class of generalized Newtonian fluids (GNFs), also known as viscous
inelastic fluids, that exhibit shear-thinning effects, as for the blood \cite{cherry2013shear, gijsen1999influence}.
In this case, the actual shear stress depends on the shear rate at the current
time and not upon the history of the deformation rate \cite{poole2023inelastic},
and it can be described via a generalized form of the constitutive equation of
Newtonian fluids \cite{bird1987dynamics}, in which the effective (or apparent) viscosity
is a non-linear decreasing function of the shear rate.
Specifically, such a nonlinear viscosity response generally exhibits
small decreasing rates (thereby corresponding to a quasi-Newtonian behaviour)
for both low and high shear rate levels, with a significant nonlinear transition
at intermediate shear rates \cite{bird1987dynamics, chhabra2011non}.

Several empirical models can be found in the specialized literature to describe
the rheological behaviour of GNFs with a shear-thinning response. The power-law
(or Ostwald–de Waele) model \cite{waele1923viscometry,ostwald1925ueber}, defined
by two model parameters, expresses the dependence of shear stress on shear rate
via a power law. It generally provides a suitable description of the nonlinear
viscosity behaviour in flow regimes characterized by intermediate shear rates.
The Carreau model based on molecular theoretical considerations \cite{carreau1972rheological}
and the Ellis model \cite{matsuhisa1965analytical, bird1987dynamics}, are both
formulated by introducing three model parameters. They capture the transition
from the quasi-Newtonian regime at low shear rates to the highly nonlinear
trend as the shear rate increases.
The Cross model \cite{cross1965rheology}, defined by four model parameters,
allows to capture the transitions between the quasi-Newtonian regime and the
non-Newtonian one at both low and high shear rates.
The Carreau-Yasuda model \cite{bird1987dynamics, yasuda1979investigation}, also
known as the Bird-Cross-Carreau-Yasuda (BCCY) model \cite{gallagher2019non},
aims to provide a comprehensive description of the shear-thinning behaviour
by combining both the Cross and Carreau models and introducing the
additional Yasuda parameter \cite{yasuda1979investigation}.

Specifically addressing the BCCY model, it is generally considered to be effective
and highly versatile in describing the shear-thinning trend of complex fluids.
It has been widely adopted in many advanced inherent applications, for instance,
hemodynamics \cite{gijsen1999influence}, plastic manufacturing \cite{mazzanti2016rheological},
lubricant production \cite{bair2006more}, and food processing \cite{meza2021effect}.
However, as highlighted by Gallagher et al. \cite{gallagher2019non} in the context
of the dynamics of the hematologic fluid, the BCCY model exhibits intrinsic
limitations related to the identifiability of the parameters.
The identifiability is defined as the capability to find a
unique set of model parameters from experimental data gathered from a real system \cite{godfrey1985identifiability, bellman1970structural, guillaume2019introductory, raue2009structural}.
This property is a critical aspect of the modelling process when model parameters
are related to physical features of the analysed system (i.e., the fluid rheological
properties) \cite{godfrey1985identifiability, guillaume2019introductory, raue2009structural}.
Specifically, Gallagher et al. \cite{gallagher2019non}
found that fitting experimental data using standard nonlinear regression
procedures can yield multiple sets of model parameters that provide nearly identical
fits of rheological data. This is due to the presence of large, flat regions
in the cost surface around the optimal state. Despite producing almost
indistinguishable rheological fits, these parameter variations can lead to significantly
different and unreliable flow predictions, making it impossible to draw meaningful
conclusions about the physical properties of the fluid. Furthermore, when the BCCY
model is used to calibrate piecewise approximations —based on the assumption of
a sharp viscosity transition between ideal Newtonian regions (at low and high
shear rates) and a power-law behaviour at intermediate shear rates,
which could be used to derive analytical solutions \cite{santesarti2024quasi}—
additional inconsistencies in parameter interpretation can arise, as it will be
analysed in the following. These inconsistencies compromise the physical meaning
and applicability of the estimated parameters, further limiting the robustness
and reliability of the BCCY model in such scenarios.

In order to contribute to overcoming these identifiability issues, this paper
proposes a novel heuristic rheological description for shear-thinning inelastic
fluids.
Comparisons based on analytical closed-form assessments and numerical
optimization procedures -considering both direct and inverse identification
strategies- highlight the accuracy and reliability of the proposed rheological
model.
In particular, the results emphasize the clear physical significance of
the model parameters and specifically of the shear-thinning index
that predominantly influence the nonlinear response at
intermediate shear rate levels.
Moreover, they demonstrate that the proposed
formulation is highly accurate and efficient even with a simple direct
parameter estimation from experimental data,
without requiring nonlinear regression procedures.
Furthermore, unlike classical approaches that typically refer to a single Carreau
number \cite{tabakova2020oscillatory,shahsavari2015mobility}, the proposed
formulation naturally allows for the identification of two distinct Carreau numbers,
each associated with one of the fluid's characteristic shear rates.

The paper is organized as follows. Section \ref{sec:BCCY_model} presents a
critical analysis of the BCCY formulation. In Section \ref{sec:SRB_model}, the
novel shear-based rheological model is introduced, with analytical developments
(complemented by \ref{appendixA}) providing theoretical insights into the improved
physical interpretability of its parameters.
Section \ref{sec:iden props} discusses identifiability performance and numerical
comparisons between the proposed description and the BCCY model, presenting
results of direct and inverse parameter estimation
procedures based on available experimental data, and analysing their impact on a
representative flow case study. Finally, concluding remarks are provided in
Section \ref{sec:conclusions}.

\section{An insight into the Carreau-Yasuda model}
\label{sec:BCCY_model}
  
For incompressible GNFs, the constitutive relationship between the deviatoric
stress tensor $\vectb{\tau}$ and the strain-rate tensor $\vectb{E}$  results in \cite{bird1987dynamics}
\begin{equation}
    \vectb{\tau} \left(\dot{\gamma}\right) =
    2\mu\left(\dot{\gamma}\right) \vectb{E}\ = \mu \left(\dot{\gamma}\right)\  \left( \nabla\vectb{v}+\nabla^T\vectb{v} \right)\,,
    \label{eq:GNF fluids}
\end{equation}
where  $\vectb{v}$ is the fluid velocity, $\nabla$ denotes the gradient operator,
and $\mu\left(\dot{\gamma}\right)$ is the effective viscosity depending on
the scalar measure $\dot{\gamma}$ of the strain-rate tensor
\begin{equation}\label{eq:strain rate magn}
   \dot{\gamma}=\left|2\vectb{E}\ \right|=\sqrt{2\,\text{tr}\left( \vectb{E}^T\vectb{E} \right)}=\sqrt{2\,I_2}\,,
\end{equation}
with $I_2$ being the second principal trace of the infinitesimal strain-rate tensor \cite{itskov2007tensor, bird1987dynamics, irgens2014rheology}.
Regarding GNFs that exhibit a shear-thinning behaviour and addressing the
5-parameters BCCY rheological formulation, the effective viscosity in
Eq. \eqref{eq:GNF fluids} may be described as \cite{bird1987dynamics, yasuda1979investigation}
\begin{align}
  \mu \left( \dot{\gamma} \right)  = \mu_\infty + \frac{\mu_0 - \mu_\infty}{ \left[ 1+ \left( \lambda\dot{\gamma} \right)^a \right]^{\frac{1-n}{a}} }\, ,
  \label{eq:BCCY_model}
\end{align}
where $\mu_0$ and $\mu_\infty$ (with $\mu_\infty < \mu_0$) are the zero-shear rate
viscosity and the infinity-shear rate viscosity values (measured in $[\text{Pa}\cdot \text{s}]$)
attained for very low  and very high shear rate levels, respectively, $\lambda$
(measured in $[\text{s}]$)
is a characteristic timescale inversely proportional to the  shear rate
threshold marking the onset of
significant shear-thinning effects \cite{shahsavari2015mobility, poole2023inelastic},
$a$ is a dimensionless strictly-positive model
parameter (generally greater than 1), also referred to as the Yasuda parameter
\cite{yasuda1979investigation} and which adjusts the transition between quasi-Newtonian
regimes and the dominant shear-thinning response, and $n$ is a dimensionless index
regulating the corresponding shear-thinning rate, such that $n\in (0,1)$.

\begin{figure}[tb]
    \centering
    \includegraphics[width=0.8\textwidth]{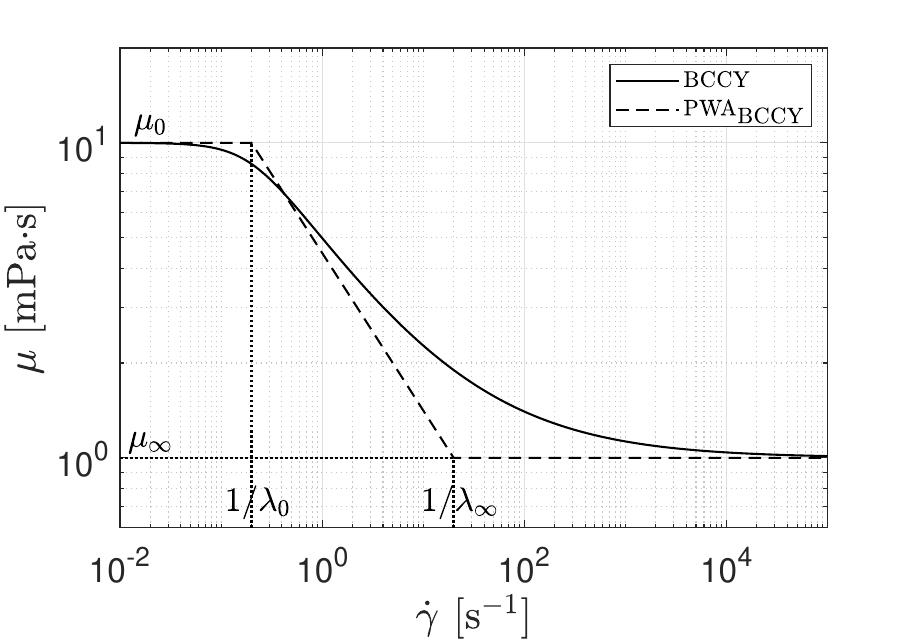}
    \caption{Example case of the Bird-Cross-Carreau-Yasuda (BCCY) shear-thinning description in comparison with a power-law-based piecewise approximation defined in terms of the BCCY model parameters (PWA$_\text{BCCY}$), i.e. by considering $N=n$, $\lambda_0=\lambda$, $\lambda_\infty=\lambda(\mu_\infty/\mu_0)^{1/(1-n)}$, $K= \mu_0 \lambda^{n-1}$. Values of model parameters: $\mu_0 = 10\, \text{mPa}\cdot \text{s}$, $\mu_\infty = 1\, \text{mPa}\cdot \text{s}$, $\lambda = 5\,  \text{s}$, $n = 0.5$, $a=2.0$.}
    \label{fig:BCCY_PWA_ex}
\end{figure}

Referring to the example case shown in Fig. \ref{fig:BCCY_PWA_ex}, the behaviour
captured by the BCCY model is quantitatively consistent with typical experimental
observations for many GNFs exhibiting shear-thinning responses \cite{gijsen1999influence, bair2006more, zare2019analysis, ohta2005dynamic}
and is characterized by:
\begin{itemize}
    \item a quasi-Newtonian regime at low shear rates (namely, for $\dot{\gamma} \ll 1/\lambda$), associated to a quasi-constant viscosity value $\mu \simeq \mu_0$;

    \item a strongly nonlinear shear-thinning response for intermediate shear
    rates (i.e., for $\dot{\gamma} > 1/\lambda$);

    \item a quasi-Newtonian regime at high shear rates (namely, for $\dot{\gamma} \gg 1/\lambda$), associated to the asymptotic viscosity value $\mu_\infty$.
\end{itemize}

The range of shear rates where the fluid experiences a significant shear-thinning
response  is usually addressed as power-law region, since it can be effectively
described by means of the Ostwald–de Waele power-law relationship \cite{waele1923viscometry, ostwald1925ueber}.
In this case, it is possible to assume that $\tau=K\dot{\gamma}^N$, where $\tau$
is a scalar measure of the shear stress, $K$ (measured in $[\text{Pa}\cdot \text{s}^N]$)
is the so-called consistency index (with $K>0$), and $N$ is the dimensionless
power-law index. Accordingly, the effective viscosity can be expressed as
$\mu(\dot{\gamma})=\tau/\dot{\gamma}=K\dot{\gamma}^{N-1}$, corresponding to a
straight line representation in a log-log graph: $\log_{10}K $ being the intercept
of this line with the two axes, and the power-law index $N$ defining the line slope,
equal to $N-1$. Consequently, a shear-thinning response is recovered by
prescribing $N\in (0,1)$. If  outside from the power-law region the fluid is
approximated as perfectly Newtonian, the viscosity dependency on the shear rate
may be described via a power-law-based piecewise approximation (PWA), resulting in
\begin{align} 
    \mu \left(\dot{\gamma}\right) =
    \begin{cases}
        \mu_0               & \text{for $\dot{\gamma} \le  1/{\lambda}_0$} \\
        K\dot{\gamma}^{N-1} & \text{for $1/{\lambda}_0 \le \dot{\gamma} \le 1/{\lambda}_\infty$} \\
        \mu_\infty          & \text{for $\dot{\gamma} \ge 1/{\lambda}_\infty$}
    \end{cases}\, ,
    \label{eq:PWA}
\end{align}
where the continuous but sharp transitions between ideal Newtonian regimes and
the shear-thinning region are identified by the time constants $\lambda_\infty$
and $ \lambda_0$ (with $\lambda_\infty < \lambda_0$). It is worth observing
that the definition of a PWA needs the setting of four model parameters
($\mu_0, \lambda_0, N, \lambda_\infty$) that, since the continuity requirement
at the characteristic shear rates $\dot{\gamma}=1/{\lambda}_0$ and $\dot{\gamma}=1/{\lambda}_\infty$, have to satisfy
the following relationships
\begin{align}
    K=\mu_0\lambda_0^{N-1}=\mu_\infty\lambda_\infty^{N-1}\,,\qquad \frac{\mu_0}{\mu_\infty}=\left(\frac{\lambda_\infty}{\lambda_0}\right)^{N-1}\, .
    \label{eq:K_muratio_PWA}
\end{align}

A possible point of confusion might be how to introduce a PWA of the  BCCY model.
To this end, assuming the same values for $\mu_0$ and $\mu_\infty$ in both the
descriptions, as well as $\lambda_0 = \lambda$, can be clearly considered as
consistent choices. Accordingly, for a given value of $N$, the consistency index
$K$ and the time constant ${\lambda}_\infty$ directly follow from Eqs. \eqref{eq:K_muratio_PWA}.
But what about the value for the power-law index $N$? A possible direct choice
might be to prescribe that the power-law index $N$ coincides with the BCCY model
parameter $n$ \cite{pratumwal2017whole}.
Such an assumption is often implicitly enforced as a consequence of (or justifying)
the usually-adopted misleading notation, where the same symbol is generally used
to indicate both $N$ and $n$ \cite{gallagher2019non, bird1987dynamics, irgens2014rheology, chhabra2011non, tabakova2020oscillatory}.
In the following, the PWA defined by enforcing $N=n$ will be denoted as PWA$_\text{BCCY}$.

Referring to the example case introduced in Fig. \ref{fig:BCCY_PWA_ex}, the
comparison between the behaviour predicted by the BCCY model and the corresponding
PWA$_\text{BCCY}$ highlights that:
\begin{itemize}
    \item
    the PWA$_\text{BCCY}$ and the BCCY descriptions are in good agreement in
    representing the onset of the nonlinear shear-thinning response at
    $\dot{\gamma}\simeq 1/\lambda$;

    \item
    for $\dot{\gamma}> 1/\lambda$, the PWA$_\text{BCCY}$ exhibits a slope that
    significantly differs from the local BCCY slope across much of the power-law
    region;

    \item
    the nonlinear shear-thinning regime described by the BCCY model extends well
    beyond $\dot{\gamma}=1/\lambda_\infty$, where the PWA$_\text{BCCY}$ reaches
    the lower-bound viscosity value $\mu_\infty$.
\end{itemize}

The discrepancies revealed between BCCY and PWA$_\text{BCCY}$ stem from the fact
that the BCCY model parameter $n$ in Eq. \eqref{eq:BCCY_model} does not directly
correspond to the power-law index $N$ used in the power-law description.
Specifically, denoting by $n^\prime$ the slope of the shear stress versus the
shear rate in a log–log plot (namely, $n’=d(\log_{10}\tau)/d(\log_{10}\dot{\gamma})$)
and focusing on the shear-thinning region, it is evident that $n^\prime_\text{PWA}$
remains constant and equals $N$, while $n^\prime_\text{BCCY}$ varies with the
shear rate and cannot be directly expressed solely in terms of $n$.
Consequently, when experimental data are fitted by using the BCCY model, the
fitting value for $n$ should not be considered as representative of the power-law
index $N$. Similarly, if the experimental shear-thinning regime is fitted by
using the Ostwald–de Waele power-law relationship, the BCCY model defined by
assuming  $n=N$ may not generally provide a good fit, as specifically addressed
in Section \ref{sec:iden props}.

This observation is formally supported by calculating the local slope
$\mathcal{S}_\text{BCCY} = (n'_{\text{BCCY}}-1)$ of the BCCY viscosity model in a
log-log representation and normalized with respect to the power-law viscosity slope $(N-1)$.
In detail, by referring to Eq. \eqref{eq:BCCY_model}, it results in
\begin{align}
    \mathcal{S}_\text{BCCY} =\frac{n'_{\text{BCCY}}-1}{N-1} &=\frac{1}{N-1}\,\frac{d(\log_{10}\mu)}{d(\log_{10}\dot{\gamma})} \nonumber\\
    &=\frac{n-1}{N-1}\,\frac{(\mu_0-\mu_\infty)(\lambda\dot{\gamma})^a}{\mu_\infty \left[1+(\lambda\dot{\gamma})^a \right]^{1+\beta}+(\mu_0-\mu_\infty)\left[1+(\lambda\dot{\gamma})^a \right]}\ ,
    \label{eq:ratioBCCY}
\end{align}
where $\beta=\left(\frac{1-n}{a}\right)\in (0,1)$. It is simple to prove that
Eq. \eqref{eq:ratioBCCY} specialized to the case $n=N$ implies the following
inequality (see \ref{appendixA})
\begin{align}
     0<\left.\mathcal{S}_\text{BCCY}\right|_{n=N}< \frac{1-\mu_\infty/\mu_0}{1+\mu_\infty/\mu_0 \left[(1+\beta)^{1+\beta}/\beta^\beta -1\right]}\equiv \mathcal{U}< 1 \ .
     \label{eq:upperbound}
\end{align}

\begin{figure}[tb]
    \centering
    \subfloat[]{\includegraphics[width=0.6\textwidth]{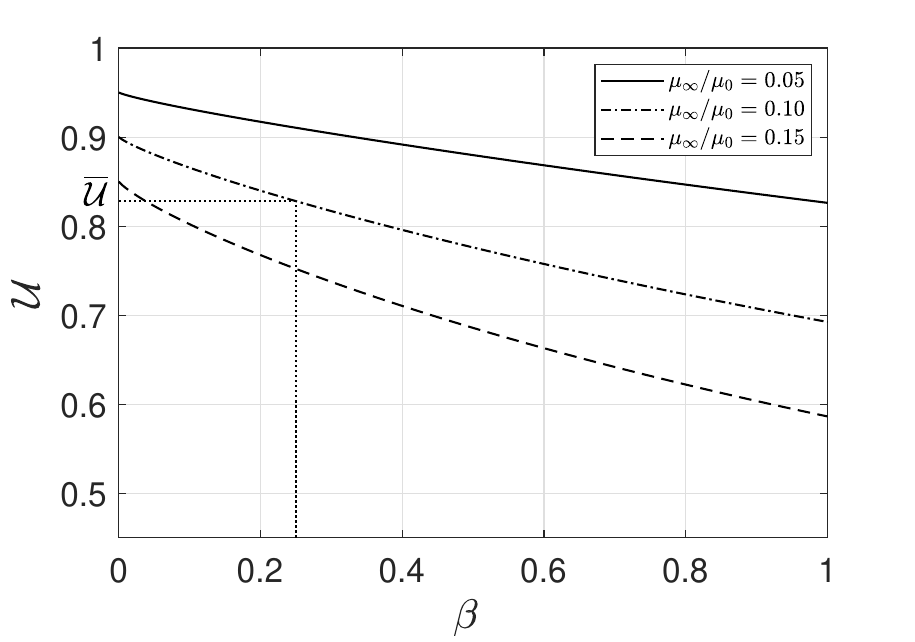}}
    \subfloat[]{\includegraphics[width=0.6\textwidth]{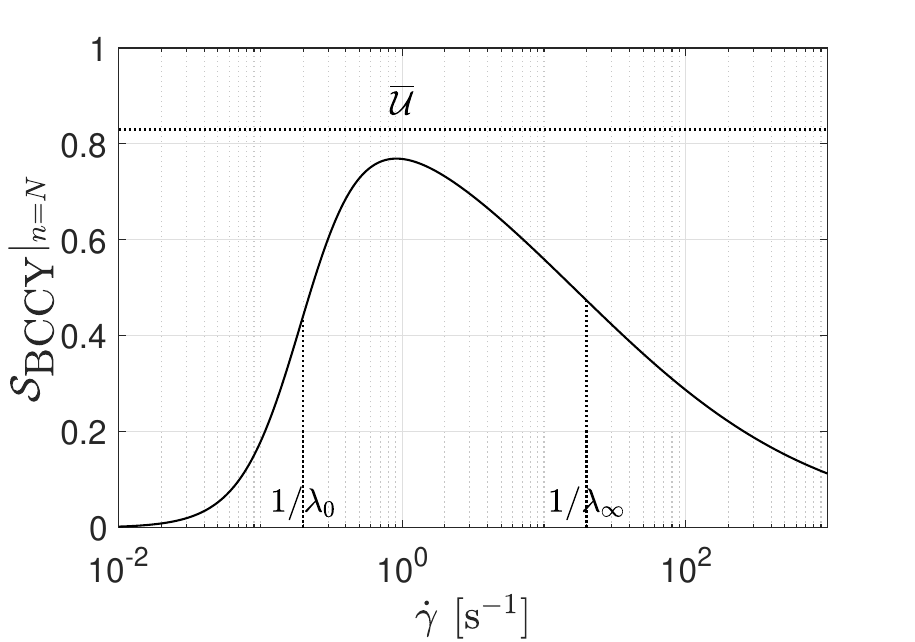}}
    \caption{Case $n=N$. (a) Upper bound $\mathcal{U}$ introduced in Eq. \eqref{eq:upperbound}
    versus $\beta=(1-n)/a$ and for different values of the ratio $\mu_\infty/\mu_0$.
    The quantity $\overline{\mathcal{U}}$ identifies the value of $\mathcal{U}$
    corresponding to the example case introduced in Fig. \ref{fig:BCCY_PWA_ex}
    (i.e., for $\beta=0.25$ and $\mu_\infty/\mu_0 =0.1$). (b) Shear rate dependency
    of the normalized BCCY slope $\mathcal{S}_\text{BCCY}|_{n=N}$ for the example
    case in Fig. \ref{fig:BCCY_PWA_ex}.}
    \label{fig:pendenzaBCCY}
\end{figure}

As a result, the dimensionless upper bound $\mathcal{U}$ decreases as the ratio
$\mu_\infty/\mu_0$ increases, becoming significantly smaller than one within the
admissible range of $\beta$.  This is clearly illustrated in Fig. \ref{fig:pendenzaBCCY},
where $\mathcal{U}$ is plotted against $\beta$ for different values of the ratio
$\mu_\infty/\mu_0$. For completeness and referring to the example case addressed
in Fig. \ref{fig:BCCY_PWA_ex}, the normalized BCCY slope $\mathcal{S}_\text{BCCY}|_{N=n}$
introduced in Eq. \eqref{eq:ratioBCCY} is also shown, revealing that the discrepancy
between the local BCCY slope and the power-law one is always greater than $23\%$,
resulting $\mathcal{S}_\text{BCCY}|_{n=N}\le 0.77< \overline{\mathcal{U}}\simeq 0.83$,
$\overline{\mathcal{U}}$ being the corresponding value of $\mathcal{U}$.

Therefore, the local slope of the BCCY-based behaviour in a log-log representation
does not accurately represent the slope of a power-law description based on the
BCCY model parameters, except in the limiting case $\mu_\infty/\mu_0\rightarrow 0^+$.
In particular, the indices $n$ in Eq. \eqref{eq:BCCY_model} and $N$ in Eq. \eqref{eq:PWA}
have different meanings, and assuming the same value for them is not consistent.
This issue can lead to a misleading physical interpretation of results obtained
using the BCCY model, and may pose a significant challenge for BCCY model parameter
identifiability when combined with power-law approximations.

\section{A novel shear-based rheological description}
\label{sec:SRB_model}

In order to overcome the issues associated to the BCCY model and discussed before,
a novel heuristic rheological description is herein proposed.

The rationale arises from the observation that the viscosity function $\mu(\dot{\gamma})$
in GNFs can be thought as the transfer function between the shear stress scalar measure $\tau$
and the applied shear rate $\dot{\gamma}$,  the corresponding PWA being the asymptotic
description of the Bode-like  diagram. Such a transfer function, in agreement with both
the previously-recalled physical evidence for shear-thinning inelastic fluids and
basic concepts of the control theory, has to be characterized by the static gain $\mu_0$,
and by the presence of both a pole and a zero, each with a multiplicity of one.
Specifically, the pole corresponds to the characteristic shear rate level $\dot{\gamma}_0=1/\lambda_0$,
which marks the decrease onset of $\mu$ from $\mu_0$; the zero is associated to
the characteristic shear rate level $\dot{\gamma}_\infty = 1/\lambda_\infty$
beyond which $\mu$ stabilizes towards the constant value $\mu_\infty$.
Accordingly, the following rheological model is proposed:
\begin{align}
\label{eq:SRB_model}
    \mu \left(\dot{\gamma}\right)=\mu_0\left[\frac{1+\left(\lambda_\infty\dot{\gamma}\right)^\alpha}{1+\left(\lambda_0\dot{\gamma}\right)^\alpha}\right]^\frac{1-\eta}{\alpha} \, .
\end{align}

This description involves five parameters: $\mu_0$, $\lambda_0$, $\lambda_\infty$,
and the positive dimensionless quantities $\alpha$ and $\eta$, which have meanings
formally equivalent to those of $a$ and $n$, respectively, in Eq. \eqref{eq:BCCY_model}.
Thereby, when $\lambda_\infty < \lambda_0$ and $\alpha>1$, a shear-thinning behaviour
is obtained if $\eta\in (0,1)$.

It is worth noticing that, with respect to the BCCY model, the rheological
description in Eq. \eqref{eq:SRB_model} emphasizes the role of the time constant
$\lambda_\infty$ as an independent parameter instead of $\mu_\infty$, this latter
simply resulting for $\dot{\gamma}\rightarrow+\infty$ from
\begin{align}
   \mu_\infty=\mu_0\left(\frac{ \lambda_\infty}{\lambda_0}\right)^{1-\eta} \, .
   \label{eq:SRB_muinf}
\end{align} 

In other words, dominant features of the shear-thinning response are assumed to
be associated to the characteristic shear rate levels $\dot{\gamma}_0 = 1/\lambda_0$ and
$\dot{\gamma}_\infty = 1/\lambda_\infty$, that identify the shear rate range in
which the highly non-Newtonian response occurs, rather than the asymptotic viscosity
value associated to the quasi-Newtonian regime. For this reason the rheological
description in Eq. \eqref{eq:SRB_model} will be denoted in the following as shear
rate-based model (SRB).

The proposed model is able to recover two other possible rheological shear-thinning
responses:
\begin{itemize}
    \item
    In the limit for $\lambda_\infty\rightarrow 0^+$, Eq. \eqref{eq:SRB_model}
    reduces to
    \begin{align}
      \mu\left(\dot{\gamma}\right)=\frac{\mu_0}{\left[1+\left(\lambda_0\dot{\gamma}\right)^\alpha\right]^\frac{1-\eta}{\alpha}}
      \label{eq:SRB_sub_model_1}\, ,
    \end{align}
    that corresponds to the Yasuda model \cite{yasuda1979investigation}, or
    equivalently to the limit condition associated to the BCCY model for
    $\mu_\infty\rightarrow 0^+$. This response is representative of a quasi-Newtonian
    regime characterized by $\mu\simeq \mu_0$ for small shear rates
    (i.e., for $\dot{\gamma}<1/\lambda_0$), followed by an indefinite shear-thinning
    behaviour for any value of the shear rate greater than $1/\lambda_0$.

    \item
    In the limit for $ \lambda_0 \rightarrow +\infty$, and accounting for
    Eq. \eqref{eq:SRB_muinf}, Eq. \eqref{eq:SRB_model} reduces to
    \begin{align}
    \mu\left(\dot{\gamma}\right) = \frac{K}{{\dot{\gamma}}^{1-\eta}}\left[1+\left(\lambda_\infty\dot{\gamma}\right)^\alpha\right]^\frac{\left(1-\eta\right)}{\alpha} \, ,
      \label{eq:SRB_sub_model_2}
    \end{align}
    with $K=\mu_\infty\lambda_\infty^{\eta-1}$, which describes a shear-thinning
    behaviour for shear rates lower than $1/\lambda_\infty$ (with a viscosity
    singularity at $\dot{\gamma}=0$), followed by a quasi-Newtonian regime
    characterized by $\mu\simeq\mu_\infty$ for $\dot{\gamma}>1/\lambda_\infty$.
\end{itemize}

It is important to emphasize that the model parameter $\eta$ introduced in
Eq. \eqref{eq:SRB_model} furnishes, in comparison with $n$ in the BCCY description,
a more effective representation of the power-lax index $N$ associated with a
SRB-based PWA. As a matter of fact, the local slope $\mathcal{S}_\text{SRB}=(n'_{\text{SRB}}-1)$
of the SRB viscosity model in a log-log representation
and normalized with respect to the power-law viscosity slope $(N-1)$, is
\begin{align}
    \mathcal{S}_\text{SRB} =\frac{n'_{\text{SRB}}-1}{N-1} &=\frac{1}{N-1}\,\frac{d(\log_{10}\mu)}{d(\log_{10}\dot{ \gamma})}\nonumber\\
    &=\frac{\eta-1}{N-1}\,\frac{\dot{\gamma}^a\left(\lambda_0^a-\lambda_\infty^a\right)}{\left[1+\left(\lambda_0\dot{\gamma}\right)^a\right]\left[1+\left(\lambda_\infty\dot{\gamma}\right)^a\right]}\ .
    \label{eq:ratioSRB}
\end{align}

In order to perform a consistent comparison among the BCCY, PWA and SRB descriptions,
from now on $\mu_0$ and $\mu_\infty$ are assumed to have the same values in all
the models, and the other parameters are assumed to satisfy the following consistency
conditions (see Eq. \eqref{eq:SRB_muinf}): $\alpha=a$,  $\lambda_0=\lambda$,
$\lambda_\infty=\lambda_0(\mu_\infty/\mu_0)^{1/(1-\eta)}$, $\eta=n=N$. In this case,
it can be shown that (see \ref{appendixA})
\begin{align}
    &0<\mathcal{S}_\text{BCCY}<\mathcal{S}_\text{SRB}<(\mathcal{S}_\text{SRB})_{max}<1 \qquad \text{for }
    \,\, \frac{1}{\lambda_0}<\dot{\gamma}<\frac{1}{\lambda_\infty} \ ,
\label{eq:DSg0}
\end{align}
where 
\begin{align}
    (\mathcal{S}_\text{SRB})_{max}=\frac{\lambda_0^{a/2}-\lambda_\infty^{a/2}}{\lambda_0^{a/2}+\lambda_\infty^{a/2}}
    \label{eq:SSRBmax}
\end{align}
is the maximum value of the shear rate function $\mathcal{S}_\text{SRB}$  attained
at $\dot{\gamma}_{max}=(\lambda_0 \lambda_\infty)^{-a/2}\in (1/\lambda_0,1/\lambda_\infty)$.

\begin{figure}[tb]
    \centering
    \subfloat[]{\includegraphics[width=0.6\textwidth]{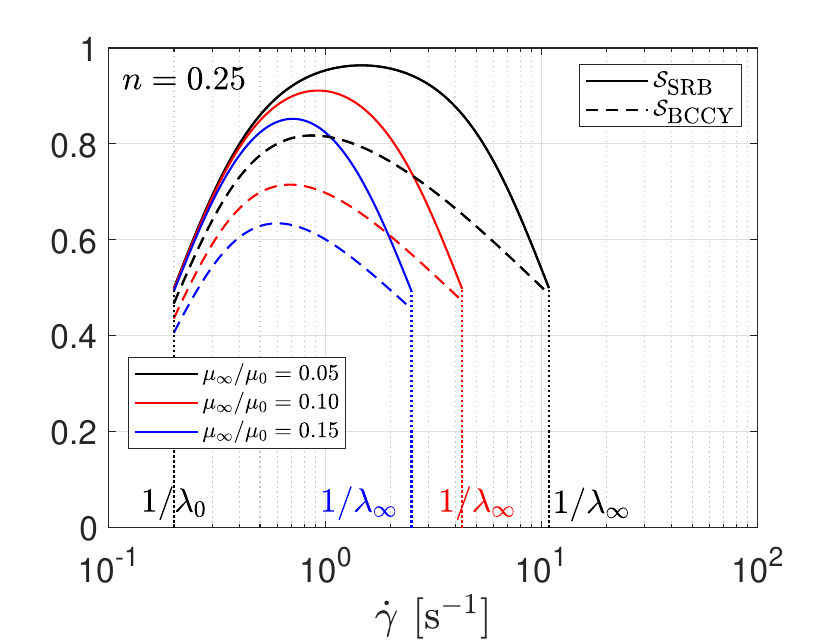}\hspace{-0.4cm}}
	\subfloat[]{\includegraphics[width=0.6\textwidth]{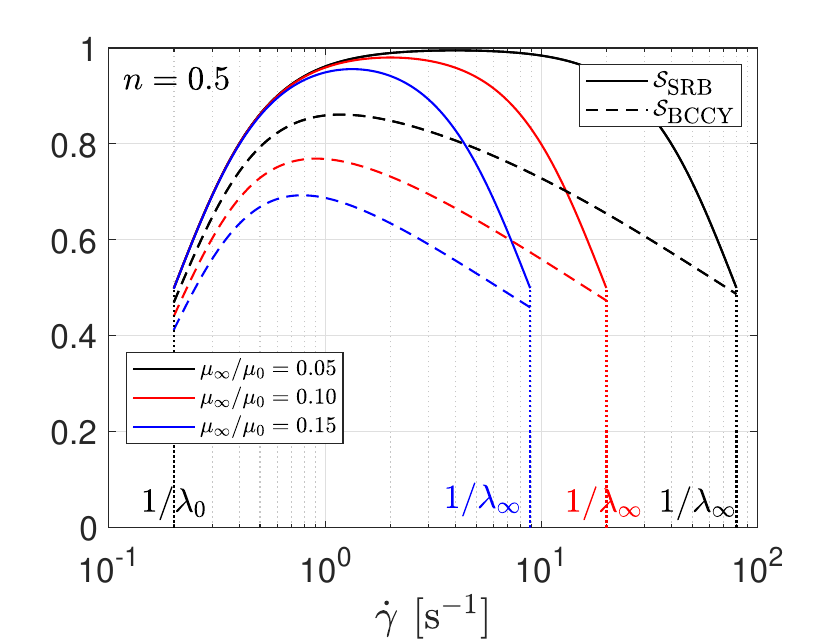}}
    \caption{shear rate dependency of local normalized slopes $\mathcal{S}_\text{BCCY}$
    (dashed lines) and $\mathcal{S}_\text{SRB}$ (continuous lines) introduced
    in Eqs. \eqref{eq:ratioBCCY} and \eqref{eq:ratioSRB}, respectively.
    Values of model parameters: $\alpha=a=2$,  $\lambda_0=\lambda=5\,\text{s}$,
    $\lambda_\infty=\lambda_0(\mu_\infty/\mu_0)^{1/(1-\eta)}$, $\eta=n=N$.
    (a) Case $n=0.25$. (b) Case $n=0.5$.}
    \label{fig:pendenzaSRB}
\end{figure}
\begin{figure}[tb]
    \centering
    \includegraphics[width=0.7\textwidth]{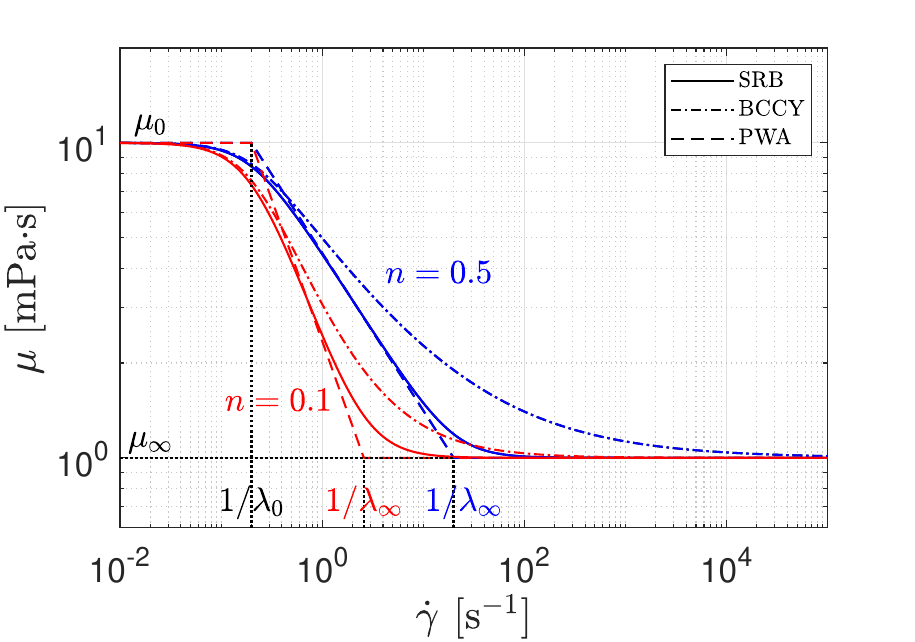}
    \caption{Shear-thinning response associated to SRB, BCCY and PWA descriptions
    and by assuming
    $\eta=n=N$, $\lambda_0=\lambda$, $\lambda_\infty=\lambda_0(\mu_\infty/\mu_0)^{1/(1-\eta)}$,
    $K= \mu_0 \lambda_0^{\eta-1}$. Values of model parameters:
    $\mu_0 = 10\, \text{mPa}\cdot \text{s}$, $\mu_\infty = 1\, \text{mPa}\cdot \text{s}$,
    $\lambda_0 = 5\,  \text{s}$, $a=2$, $n=0.5$ (blue), $n=0.1$ (red).}
    \label{fig:SRB_BCCY_PWA_ex}
\end{figure}

In this case and for different choices of model parameters, Fig. \ref{fig:pendenzaSRB}
clearly highlights that the SRB slope is significantly closer to the power-law one
within the overall shear-thinning regime than the BCCY case. Accordingly, the model
parameter $\eta$ introduced in the SRB description is expected to be much more
representative of the power-law index associated to a power-law-based approximation
than the corresponding parameter $n$ appearing in the BCCY model. This is clearly
confirmed by analysing Fig. \ref{fig:SRB_BCCY_PWA_ex} where, for different values
of $\eta=N$, the nonlinear responses experienced by considering PWA and SRB
descriptions are in a very good agreement each other, whereas the BCCY model
defined by considering $n=N$ significantly deviates from them within the
shear-thinning region.

\section{Model parameter identifiability and flow description}
\label{sec:iden props}

To highlight the parameter identifiability issues associated with the BCCY model
and to demonstrate the superior identifiability performance of the proposed SRB
description, direct and inverse estimation procedures based on available experimental
measurements are presented. Specifically, reference is made to the experimental
viscosity data obtained by Colosi et al. \cite{colosi2016microfluidic} for a cell-laden
% hydrogel, which exhibits a typical shear-thinning response.
hydrogel, which exhibits a typical
shear-thinning response, that is a decreasing shear rate dependency of viscosity.
For the sake of simplicity, the model parameters $\alpha$ (for SRB) and $a$ (for BCCY)
are assumed to be equal, and characterized by the a-priori fixed value $\alpha=a=2$
which gives accurate fits for many polymeric fluids and melts \cite{bird1987dynamics, irgens2014rheology}.
Finally, the performance of both rheological models in capturing the flow properties
of a representative shear-thinning fluid is briefly discussed.
Specifically, in order to clearly highlight the descriptive capability of the models,
independent of possible effects related to the specificity of the application,
reference is made to the simple case of a steady flow in a cylindrical channel.

\subsection{Direct parameter estimation from experimental data}
\label{sec:direct_calibration}

\begin{figure}[tb]
      \centering
         {\includegraphics[width=0.75\textwidth]{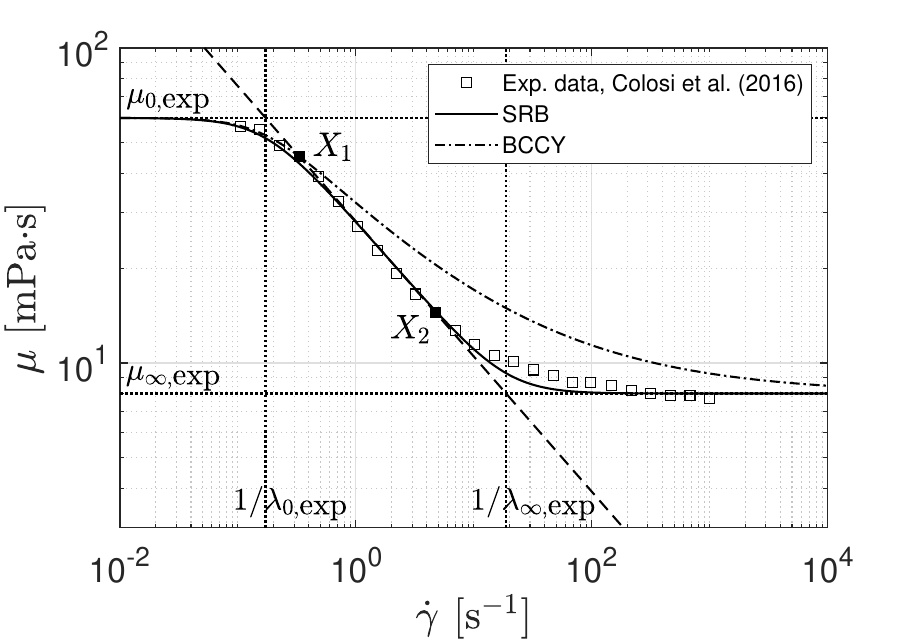}}
      \caption{Direct parameter estimation based on a straightforward heuristic visual fitting of the experimental data reported in \cite{colosi2016microfluidic}. Comparison between the corresponding SRB and BCCY rheological descriptions, defined by the experimentally-derived model parameters summarized in  Table \ref{tab:exp derived params} ($a=\alpha=2$, $n=\eta=N_{\text{exp}}$, $\lambda=\lambda_{0,\text{exp}}$). }
      	  \label{fig:from exp data to models}
\end{figure}

Parameters for both the BCCY and the SRB models are initially identified by using
a direct estimation procedure from experimental data, without employing
iterative approaches based on nonlinear regression techniques.
Specifically, Fig. \ref{fig:from exp data to models} shows the experimental
data (represented by square symbols) as represented via a log-log plot in the
plane $(\dot{\gamma},\mu)$. The adopted direct estimation procedure is
defined by the following steps:
\begin{itemize}
    \item Identification of asymptotic viscosity values. The experimental
    estimates for the asymptotic viscosity values, denoted as $\mu_{0,\text{exp}}$
    and  $\mu_{\infty,\text{exp}}$ are assumed to approximate the largest and
    smallest viscosity experimental measurements, respectively.
    In Fig. \ref{fig:from exp data to models}, $\mu_{0,\text{exp}}$ and
    $\mu_{\infty,\text{exp}}$ are represented by horizontal dot lines.
    \item Identification of the power-law region. The power-law region is
    identified by considering the portion of the experimental data that in the
    log-log plot exhibits a linear trend. In the case under consideration
    (Fig. \ref{fig:from exp data to models}), such a linear trend is described
    through the straight dashed line passing through the experimental data
    points $X_1\equiv ({\dot{\gamma}}_1,\mu_1)$ and  $X_2\equiv ({\dot{\gamma}}_2,\mu_2)$.
    Accordingly, the experimentally-deduced power-law parameters, namely the
    consistency index $K$ and the dimensionless power-law index $N$, straightly
    result from
      \begin{equation}\label{eq:n derived exp}
        K_{\text{exp}}^\ast =\frac{\mu_{1}^\ast{\dot{\gamma}}_{2}^\ast-\mu_{2}^\ast{\dot{\gamma}}_{1}^\ast}{{\dot{\gamma}}_{2}^\ast-{\dot{\gamma}}_{1}^\ast}\,,\qquad N_{\text{exp}}=1-\frac{\mu_{1}^\ast-\mu_{2}^\ast}{{\dot{\gamma}}_{2}^\ast-{\dot{\gamma}}_{1}^\ast}\,,
      \end{equation}
    where $q^\ast=\log_{10}{q}$.
    \item Determination of time constants. Since Eqs. \eqref{eq:K_muratio_PWA},
    the experimentally-based time constants $\lambda_{0,\text{exp}}$ and
    $\lambda_{\infty,\text{exp}}$ are determined as follows:
        \begin{equation}
    	    \lambda_{0,\text{exp}}=\left(\frac{K_{\text{exp}}}{\mu_{0,\text{exp}}}\right)^\frac{1}{N_{\text{exp}}-1}\,,\qquad {\lambda}_{\infty,\text{exp}}=\left(\frac{K_{\text{exp}}}{\mu_{\infty,\text{exp}}}\right)^\frac{1}{N_{\text{exp}}-1}\,.
        \end{equation}
    The inverses of these constants correspond to the characteristic shear
    rate levels at which the power-law dashed line in Fig. \ref{fig:from exp data to models}
    intersects the two horizontal dotted lines representing the asymptotic viscosity values.
\end{itemize}

\begin{table}[tb]
          \caption{Values of model parameters derived through the direct
          estimation procedure from the experimental data reported in \cite{colosi2016microfluidic}.}
          \centering % centering table
          \vspace{5pt}
          \begin{tabular}{cccccc} 
            \hline
            $\mu_{0,\text{exp}}$ & $\mu_{\infty,\text{exp}}$ & $K_\text{exp}$ & $N_\text{exp}$ & $\lambda_{0,\text{exp}}$ & 
            $\lambda_{\infty,\text{exp}}$\\
            $[\text{mPa $\cdot$ s}]$ &  $[\text{mPa $\cdot$ s}]$ & $[\text{mPa $\cdot$ s$^N$}]$ & $[-]$ & $[\text{s}]$ & $[\text{s}]$   \\
            \hline 
            60.0  & 8.0 & 28.2 & 0.571 & 5.812  & 0.053 \\
	  		\hline
         \end{tabular}
         \label{tab:exp derived params}
\end{table}

The values of model parameters derived from the experimental viscosity data through
the described direct identification procedure are reported in Table \ref{tab:exp derived params}.
Assuming $n=\eta=N_{\text{exp}}$ and $\lambda=\lambda_{0,\text{exp}}$,
Fig. \ref{fig:from exp data to models} represents a comparison between the rheological
responses obtained from the BCCY and SRB descriptions. As it clearly appears,
the SRB model demonstrates excellent fitting capability, whereas the BCCY model
provides a good fit only in the transition region at $\dot{\gamma}\simeq 1/\lambda_{0,\text{exp}}$,
failing to accurately capture the experimental behaviour in the power-law region
and the transition towards the asymptotic quasi-Newtonian response at
$\mu\simeq \mu_{\infty,\text{exp}}$.

Accordingly, the herein-described simple direct estimation procedure provides
an effective and accurate SRB-based rheological description. However, it proves
insufficient for the BCCY model, which requires an inverse identification
procedure based on nonlinear regression techniques, as discussed in the following.

\subsection{Inverse parameter identification through a nonlinear regression technique}
\label{sec:invcalibration}

Model parameter estimation is now carried out through a nonlinear regression
procedure based on the maximum likelihood estimation method \cite{aho2013foundational, motulsky2004fitting}.

Due to the nonlinearity of the models under investigation, parameter
identification may depend on both the choice of data representation and the
definition of the loss function \cite{singh2019fitting, bailer1990note, freund2015quantitative}.
In this study, to enhance fitting performance and enable meaningful model comparison,
a log–log data representation is adopted, and the lognormal mean absolute
percentage error (LMAPE) is used as the loss function in place of the
conventional sum of squared errors. Specifically,
the optimization problem aims to determine the set of model parameters that
minimizes the LMAPE, defined as:
\begin{align} 
   \text{LMAPE} = \frac{1}{T} \sum_{i=1}^{T} \left| \frac{\log{f_i}}{\log{b_i}} - 1 \right| \,,
\end{align}
where the logarithm operator is taken to base 10, $T$ is the number of experimental
measurements, $b_i$ denotes the $i$-th measured viscosity value, and $f_i$ represents
the corresponding prediction provided by the rheological model.

In addition, the coefficient of determination $R^2$ is employed as a supplementary
metric for model comparison. In the context of nonlinear regression, its definition can
be adapted to the specific case in order to provide a quantitative measure of
goodness of fit \cite{cornell1987factors, cameron1997r, motulsky2004fitting}.
In this study,
the coefficient of determination is defined as:
\begin{align}
	 R_{\text{LMAPE}}^2=1-\dfrac{\text{LMAPE}}{\dfrac{1}{T} \mathlarger{\sum}_{i=1}^T \left| \frac{\bar{y}_L}{\log{b_i}} - 1 \right|}\, ,
\end{align}
where $\bar{y}_L = (\sum_{i=1}^T \log{b_i})/T$ represents the mean logarithmic
experimental value. A value of $R_{\text{LMAPE}}^2$ close to one indicates that
the rheological model closely fits the experimental data, while a value approaching
zero suggests poor model accuracy \cite{aho2013foundational, motulsky2004fitting}.

Furthermore, to provide a more meaningful assessment of the goodness of fit
in the context of nonlinear regression, the reduced chi-square statistic $\bar{\chi}^2$
is also employed as an additional goodness-of-fit metric \cite{taylor1997introduction}.
It is defined as:
\begin{equation}
 \bar{\chi}^2 = \dfrac{\chi^2}{\text{DoF}} = \dfrac{1}{\text{DoF}}\mathlarger{\sum}_{i=1}^T\left(\dfrac{f_i - b_i}{\sigma_{i}} \right)^2 \, ,
\end{equation}
where $\text{DoF}=(T - \#\mathcal{P})$ is the number of degrees of freedom,
$\#\mathcal{P}$ denotes the number of model parameters, and $\sigma_i$ is
the standard deviation associated with the $i$-th viscosity measurement.
Specifically, a value of $\bar{\chi}^2 \approx 1$ indicates a good fit; values
significantly larger than one suggest a poor fit, while values much smaller than
one may imply potential overfitting (i.e., the model also captures random noise
in the data) \cite{jones2019beyond, laborda2021feature}. Since the experimental
data in \cite{colosi2016microfluidic} do not report values of $\sigma_i$
they are assumed, in agreement with \cite{shaw2017uncertainty, taylor1997introduction},
to be 5\% of the corresponding measured values $b_i$, for the sake of consistent
model comparison.

As a result of an iterative numerical procedure implemented in the MATLAB
environment (R2024b, MathWorks, MA, USA), Table \ref{tab:SRB PWM BCCY fittings}
summarizes the optimal parameter values computed for both the BCCY and SRB models.
It is worth noting that, in accordance with the considerations presented
in Sections \ref{sec:BCCY_model} and \ref{sec:SRB_model}, the parameter $\lambda_\infty$
does not explicitly appears in the BCCY formulation, but is instead derived
from Eqs. \eqref{eq:K_muratio_PWA} once $\mu_0$, $\lambda$, $n$, and $\mu_\infty$  are
determined. Conversely, $\lambda_\infty$ directly appears in the SRB model introduced
in Eq. \eqref{eq:SRB_model}, while $\mu_\infty$ is derived from Eq. \eqref{eq:SRB_muinf}.
Therefore,
the parameters ($\mu_0$, $\lambda$, $n$, $\mu_\infty$) for the BCCY model
and ($\mu_0$, $\lambda_0$, $\eta$, $\lambda_\infty$ ) for the SRB model
are denoted as primary parameters, while $\lambda_\infty$
for the BCCY and $\mu_\infty$ for the SRB model are referred to as derived parameters.

For completeness, Table \ref{tab:SRB PWM BCCY fittings} also reports the optimal
LMAPE values along with the corresponding values for the coefficient of
determination $R_{\text{LMAPE}}^2$ and the reduced chi-square $\bar{\chi}^2$.
 
\begin{table}[tb]
\caption{Inverse identification of model parameters for SRB and BCCY rheological
descriptions performed via a nonlinear regression procedure based on the maximum likelihood
estimation method. Primary and derived (\underline{underlined}) fitting values
($a=\alpha=2$, $n^\ast=\eta$ and $\lambda^\ast=\lambda_0$ for SRB; $n^\ast=n$
and $\lambda^\ast=\lambda$ for BCCY).}
\centering  	  	  
\begin{tabular}{lccccccc c}
\hline
 & $\mu_{0}$ & $\mu_{\infty}$ & $\lambda^\ast$ & $\lambda_\infty$ & $n^\ast$ &	LMAPE & $R_{\text{LMAPE}}^2$ &  $\bar{\chi}^2$ \\
 &  $[\text{mPa $\cdot$ s}]$ &  $[\text{mPa $\cdot$ s}]$ &  $[\text{s}]$ & $[\text{s}]$ & $[-]$ &  $[-]$ &  $[-]$ & $[-]$\\
\hline 
SRB  & 60.0 & \underline{8.14} & 5.747 & 0.053 & 0.573 & 0.0155 & 0.93 & 0.96 \\
BCCY & 60.1 & 7.70 & 3.759 & \underline{0.209} & 0.289  & 0.0034 & 0.98 & $0.09$ \\
\hline
\end{tabular}
\label{tab:SRB PWM BCCY fittings}
\end{table}
\begin{figure}[tb]
\centering
    {\includegraphics[width=0.75\textwidth]{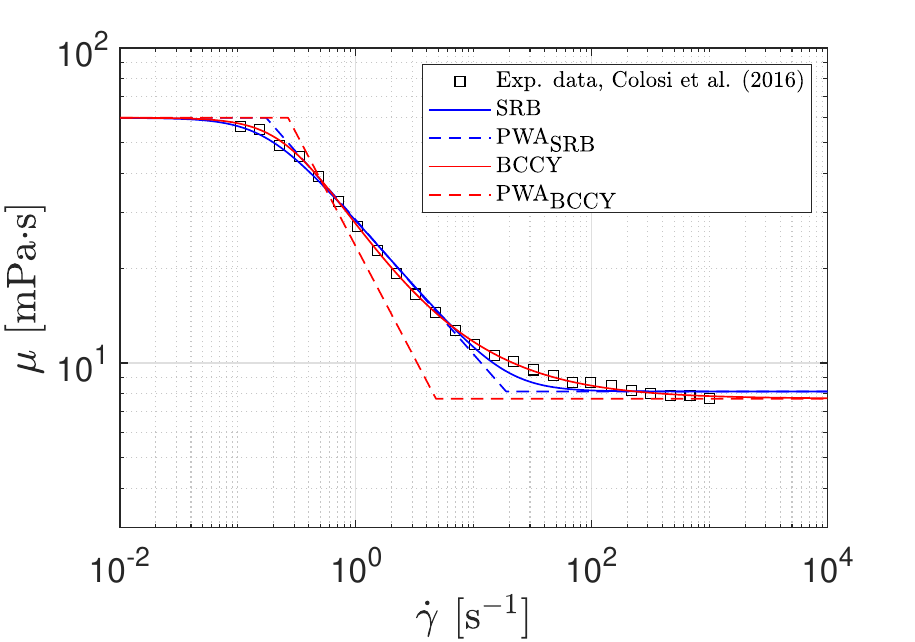}}
    \caption{Inverse identification of model parameters through numerical nonlinear regression. Comparison
    of the fitting performance of the SRB and BCCY models, along with their
    corresponding piecewise approximations (PWA$_\text{SRB}$ and PWA$_\text{BCCY}$).
    The optimal values of the model parameters are summarized in Table \ref{tab:SRB PWM BCCY fittings}.}
\label{fig:SRB PWM BCCY models to PWM}
\end{figure}
         
Specifically, the minimum LMAPE value obtained by the optimization procedure for
the SRB model is slightly higher than that of the BCCY model. Consequently, the
coefficient of determination $R_{\text{LMAPE}}^2$ for the SRB model is marginally
lower than that for the BCCY formulation.
Accordingly, this outcome suggests that, based on the computed optimal parameter
sets, the BCCY model provides a slightly better fit to the experimental data
compared to the SRB model. Nevertheless, the latter still exhibits very good
fitting performance, as clearly illustrated in Fig. \ref{fig:SRB PWM BCCY models to PWM},
where the experimental data from Colosi et al. \cite{colosi2016microfluidic}
are well fitted by both the SRB and BCCY models optimized through the numerical
nonlinear regression.

On the other hand, the computed values of the reduced chi-square $\bar{\chi}^2$
differ significantly, resulting in $\bar{\chi}^2 \approx 1$ for the SRB model
and $\bar{\chi}^2 < 0.1$ for the BCCY one. Such a result confirms the good
fitting performance of the SRB model, while also highlights a clear
tendency of the BCCY model to suffer from potential overfitting.

The comparison between the values summarized in Table \ref{tab:SRB PWM BCCY fittings}
and those reported in Table \ref{tab:exp derived params} reveals that the BCCY
optimal parameters significantly differ from those obtained through the simple
direct estimation procedure described in Section \ref{sec:direct_calibration}.
In contrast, the direct identification procedure, based on a straightforward
heuristic visual fitting approach, yields an estimation for the SRB model parameters
that closely matches the optimal one numerically obtained via the iterative
% minimization
nonlinear regression strategy.

Let the power-law-based piecewise approximations PWA$_\text{SRB}$ and PWA$_\text{BCCY}$
be considered (see Eq. \eqref{eq:PWA}), built up with the inverse identification
results associated with the SRB and BCCY descriptions, respectively. As shown in
the comparison presented in Fig. \ref{fig:SRB PWM BCCY models to PWM}, PWA$_\text{BCCY}$,
despite being based on optimal parameters that ensure the best fit for the BCCY
description, fails to effectively capture the experimental response. In contrast,
PWA$_\text{SRB}$ exhibits an excellent agreement with the experimental data,
accurately reproducing both the power-law behaviour and the transition phases
toward quasi-Newtonian regimes.

This result provides further compelling evidence of a critical identifiability
issue associated with the BCCY description, arising from the fact that the BCCY
model parameter $n$ in Eq. \eqref{eq:BCCY_model} ($n = 0.289$ at the optimality
condition) does not directly correspond to the power-law index $N$ used in the
power-law description. Instead, the proposed results confirm that the power-law
index $N$ is more accurately represented by the SRB model parameter $\eta$ in
Eq. \eqref{eq:SRB_model} ($\eta = 0.573$ at the optimality condition).

\begin{figure}[!htb]
\centering
\subfloat[]{\includegraphics[width=0.9\textwidth]{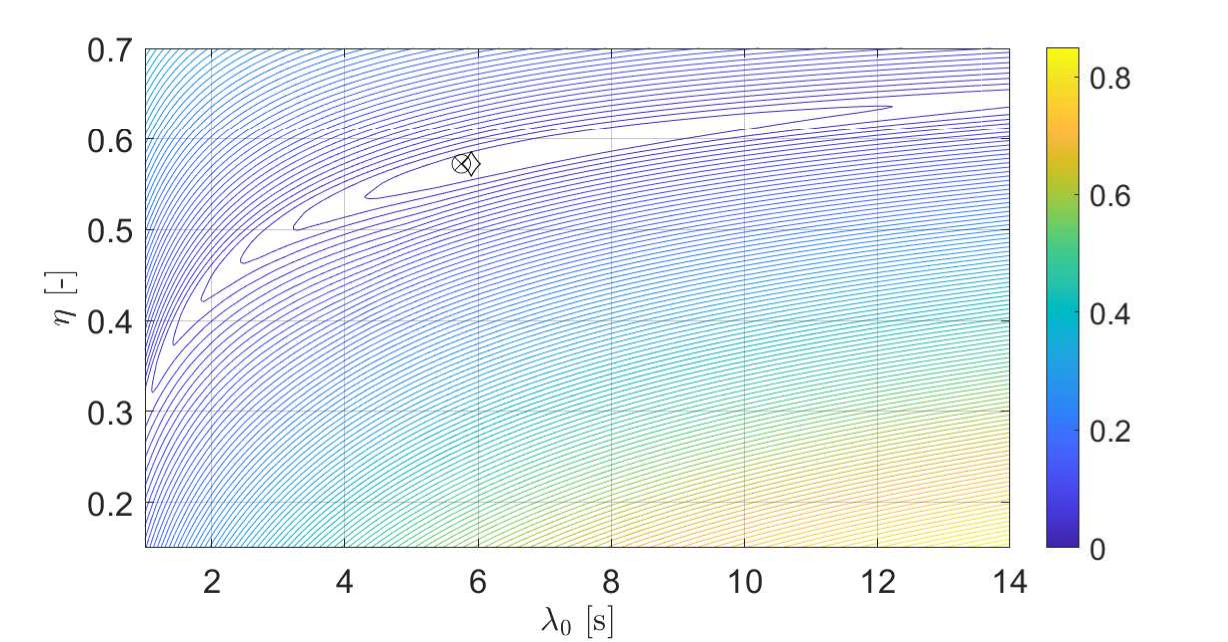}} \\
\subfloat[]{\includegraphics[width=0.9\textwidth]{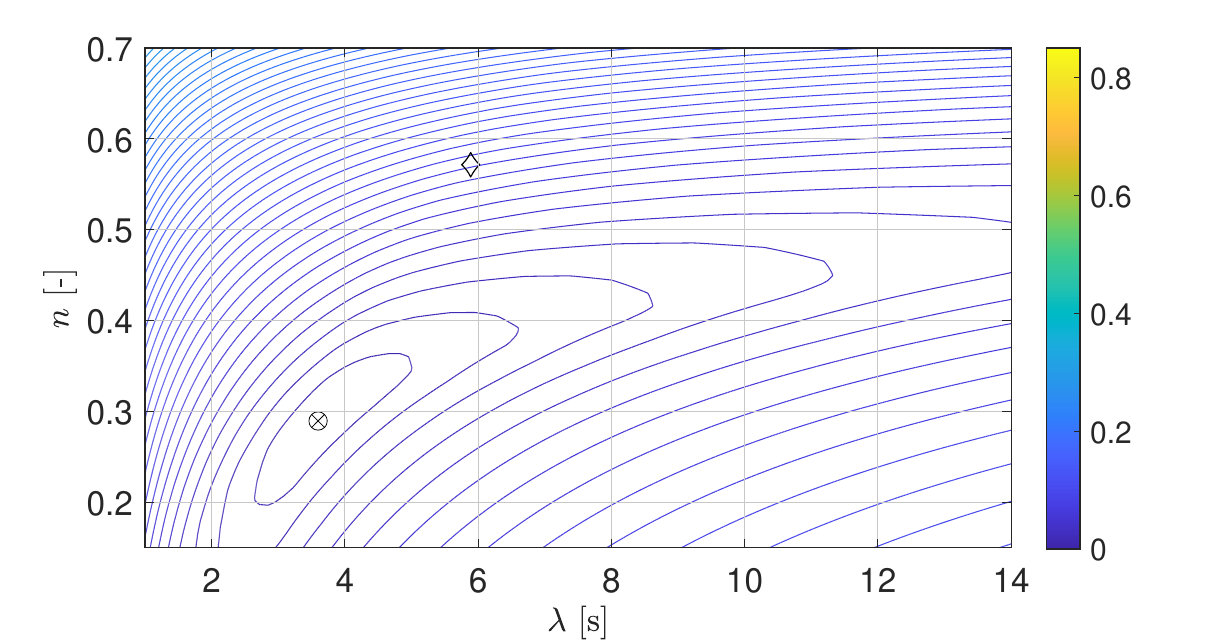}}
\caption{Isolines of the likelihood surface $\mathcal{L}(\lambda^\ast, n^\ast)$
introduced in Eq. \eqref{eq:altfits functional} ($n^\ast=\eta$ and $\lambda^\ast=\lambda_0$
for SRB; $n^\ast=n$ and $\lambda^\ast=\lambda$ for BCCY). Model parameters
different from $\lambda^\ast$ and $n^\ast$ are held fixed and equal to the optimal
values summarized in Table \ref{tab:SRB PWM BCCY fittings}.
(a) SRB model: $\mathcal{L}(\lambda_0, \eta)$;
(b) BCCY model: $\mathcal{L}(\lambda, n)$. $\otimes$: optimality condition computed
via the inverse identification strategy; $\diamondsuit$: direct estimation state.}
\label{fig:SRB BCCY altfits}
\end{figure}

Finally, to demonstrate the greater robustness of the proposed SRB model in terms
of identifiability features, particularly in comparison to the BCCY model, the
variability of LMAPE associated to deviations of model parameters from the optimality
condition is analysed.
For the sake of simplicity, reference is made to $\lambda^\ast$ and $n^\ast$ only
($n^\ast=\eta$ and $\lambda^\ast=\lambda_0$ for SRB; $n^\ast=n$ and $\lambda^\ast=\lambda$
for BCCY), as these primary parameters play a key role in capturing the onset and
evolution of the nonlinear shear-thinning response. Specifically, let the
likelihood function $\mathcal{L}$ be defined as
\begin{align}
        \mathcal{L}(\lambda^\ast, n^\ast) =  \text{LMAPE}(\lambda^\ast, n^\ast,\hat{\mathcal{P}})\,,
        \label{eq:altfits functional}
\end{align}
where $\hat{\mathcal{P}}$ denotes the set of primary model parameters complementary
to $\lambda^\ast$ and $n^\ast$, which are held fixed at their optimal values
(see Table \ref{tab:SRB PWM BCCY fittings}). Figure \ref{fig:SRB BCCY altfits}
presents the isolines of the likelihood surface $\mathcal{L}(\lambda^\ast, n^\ast)$
obtained for the SRB and BCCY descriptions,
highlighting how LMAPE varies from its optimal value when only $\lambda^\ast$
and $n^\ast$ vary.
For completeness, the result of the direct estimation procedure from
experimental data is also indicated.
Once again, it clearly appears that in the case of the SRB model the inverse and
the direct estimation procedures give very close results, differently from the
case of the BCCY description. Furthermore, the analysis of the isolines of
$\mathcal{L}(\lambda, n)$ obtained for the BCCY model (Fig. \ref{fig:SRB BCCY altfits}b)
confirms the findings provided by Gallagher et al. \cite{gallagher2019non}.
As a matter of fact, the BCCY likelihood surface exhibits a large flat region
around the optimality condition. Consequently, significant variations of model
parameters $\lambda$ and $n$ can marginally affect the fitting of the experimental
data, but they can lead to very different flow dynamics (as discussed in the following),
making these parameters unreliable for inferring the physical properties of the
investigated fluid.

By contrast, this identifiability issue is substantially mitigated in the proposed
SRB model. The corresponding likelihood surface exhibits steep gradients around
the optimal condition, indicating that even small perturbations in $\lambda_0$
and $\eta$ result in significant variations in
$\mathcal{L}(\lambda_0, \eta)$. Consequently, minor changes in the primary parameters
can noticeably affect the fit performance of the model against the experimental data.
This behaviour, combined with the fact that the SRB model parameters can be effectively
estimated through a direct approach, underscores its superior identifiability
properties, making it a more effective and robust alternative to the BCCY formulation.

\subsection{A representative case of flow description}

To further emphasize the previously stated observations regarding the impact of potential identifiability issues on the flow description of generalized Newtonian fluids, reference is made to the simple case of a steady flow in a pipe involving an exemplary shear-thinning incompressible fluid. This case can be considered representative of an extrusion-based bioprinting process \cite{chirianni2024influence, chirianni2024development, conti2022models}. In particular, a cylindrical nozzle is examined, where $R$ denotes the cross-sectional radius and $L$ the axial length (i.e., along the coordinate $z$). Accordingly, the steady flow is fully described by the following axisymmetric one-dimensional problem \cite{conti2022models, bird1987dynamics}:

\begin{align}
    & \frac{1}{r}\frac{d}{d r} \left[ r\mu(\dot{\gamma}) \frac{d v_z}{d r} \right]+\frac{\Delta p}{L}  =0 && \text{for } r\in [0,R]\nonumber\\
    &\frac{d v_z}{d r}=0 && \text{at } r=0 \label{eq:1d_flow}\\
    &v_z =0 && \text{at } r=R\nonumber
\end{align}
where $\Delta p/L=-dp/dz$ is the constant axial pressure gradient, $v_z=v_z(r)$ is the axial velocity component (the only non-zero component), $r$ is the radial coordinate, and $\dot{\gamma}$ is the scalar shear rate measure, given by $\dot{\gamma}=|d v_z/d r|$.

\begin{figure}[tb]
\centering
\includegraphics[width=0.75\textwidth]{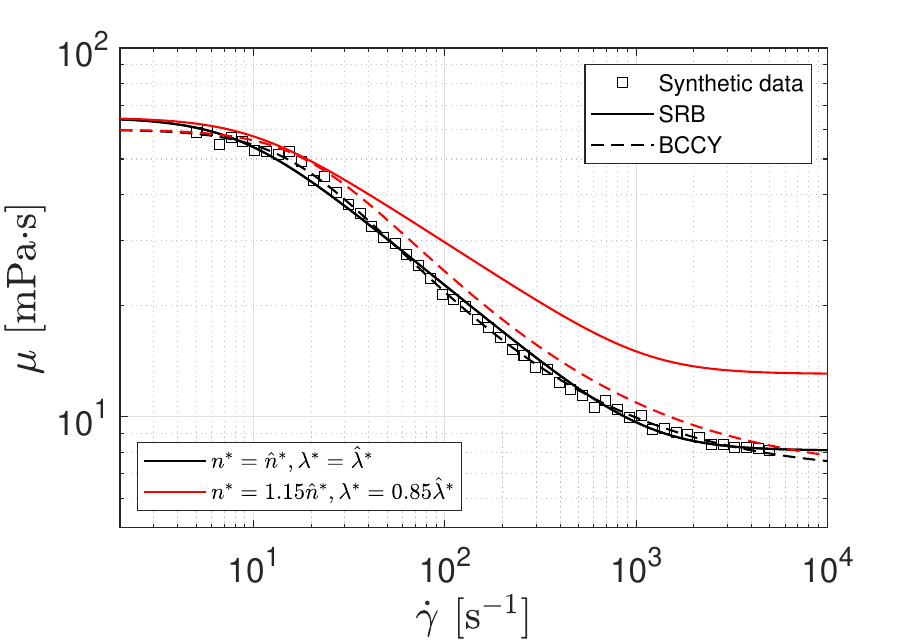}
\caption{Rheological response of a representative shear-thinning fluid defined by a set of 50 synthetic data points (square symbols). Optimal fittings (black lines) obtained through the inverse identification procedure for both the SRB (solid line) and BCCY (dashed line) formulations. The optimal model parameter values are summarized in Table \ref{tab:comparison SRB and BCCY pow law flow}. Red lines represent the rheological predictions when the primary model parameters $n^\ast$ and $\lambda^\ast$ ($n^\ast=n$ and $\lambda^\ast=\lambda$ for BCCY; $n^\ast=\eta$ and $\lambda^\ast=\lambda_0$ for SRB) are perturbed with respect to the optimal condition.}
\label{fig:synth_rheol}
\end{figure}

The rheological description of the exemplary fluid has been defined considering a set of 50 synthetic data points representing a typical shear-thinning behaviour,
and generated from the BCCY model with the addition of small lognormal noise (standard deviation 0.018), as shown in Fig. \ref{fig:synth_rheol}. The model parameters resulting from the numerical inverse identification procedure introduced in Section \ref{sec:invcalibration} are summarized in Table \ref{tab:comparison SRB and BCCY pow law flow} for both SRB and BCCY. The corresponding rheological predictions are compared in Fig. \ref{fig:synth_rheol}. Furthermore, Fig. \ref{fig:synth_rheol} illustrates the rheological predictions of both models when a variation in the primary model parameters $n^\ast$ and $\lambda^\ast$ (that is, $n$ and $\lambda$ for BCCY, and $\eta$ and $\lambda_0$ for SRB) is introduced with respect to the optimality condition. Specifically, $n^\ast$ is increased by 15\%, while $\lambda^\ast$ is reduced by 15\%, both relative to their optimal values (see Table \ref{tab:comparison SRB and BCCY pow law flow}).

\begin{table}[tb]
\caption{Primary and derived (\underline{underlined}) fitting values of model parameters, obtained through the inverse identification procedure and based on synthetic data reported in Fig. \ref{fig:synth_rheol} ($a=\alpha=2$, $n^\ast=\eta$ and $\lambda^\ast=\lambda_0$ for SRB; $n^\ast=n$ and $\lambda^\ast=\lambda$ for BCCY). }
\centering 
\begin{tabular}{lccccc}
\hline
 & $\mu_{0}$ & $\mu_{\infty}$ & $\lambda^\ast$ & $\lambda_\infty$ & $n^\ast$  \\  
 &  $[\text{mPa $\cdot$ s}]$ &  $[\text{mPa $\cdot$ s}]$ &  $[\text{s}]$ & $[\text{s}]$ & $[-]$ \\
\hline 
SRB & 60.0 & \underline{8.14} & $9.58 \cdot 10^{-2}$ & $8.86 \cdot 10^{-4}$ & 0.573\\
BCCY & 60.1 & 7.00 & $6.25 \cdot 10^{-2}$ & \underline{$4.24 \cdot 10^{-3}$} & 0.300\\
\hline
\end{tabular}
\label{tab:comparison SRB and BCCY pow law flow}
\end{table}

\begin{figure}[tb]
\centering
{\includegraphics[width=0.75\textwidth]{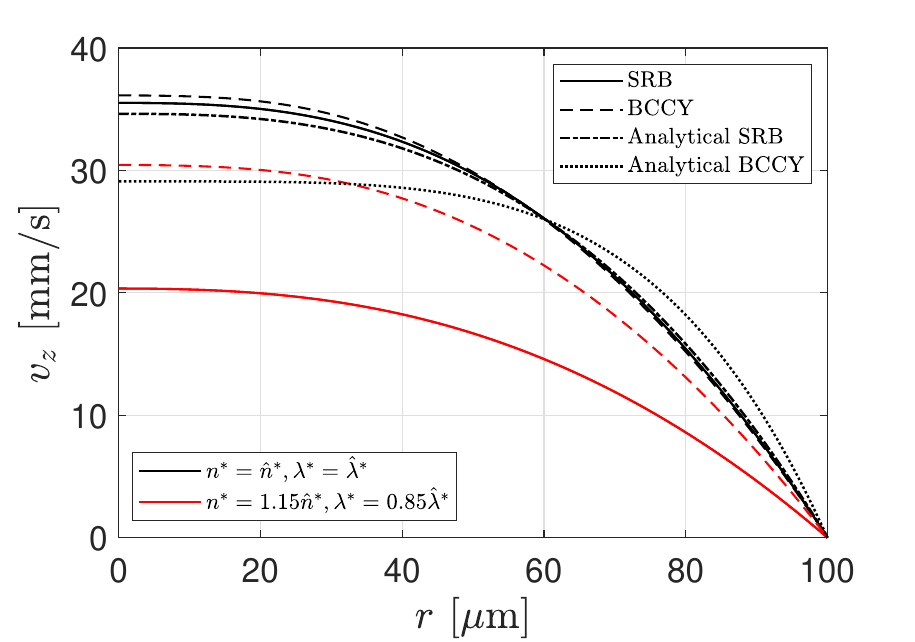}}
\caption{Prediction of velocity profiles for a steady incompressible flow in a cylindrical channel ($\Delta p/L=175.0$ Pa/mm, $R=100\, \mu$m) when the rheological shear-thinning response in Fig. \ref{fig:synth_rheol} is considered. Numerical predictions based on SRB (solid lines) and BCCY (dashed lines) models: comparison between velocity profiles associated with the optimality conditions (black lines) and the perturbed ones (red lines). Power-law-based analytical solution, described by Eq. \eqref{eq:analyticalPL} and obtained via the nonlinear regression results for both SRB and BCCY models, is also reported. The optimal model parameter values are summarized in Table \ref{tab:comparison SRB and BCCY pow law flow}.}
\label{fig:vel profiles}
\end{figure}

Starting from the fluid rheological response, the steady velocity profile can be computed by solving the problem introduced in Eqs. \eqref{eq:1d_flow}. In particular, assuming $\Delta p/L=175.0$ Pa/mm and $R=100\, \mu$m, Fig. \ref{fig:vel profiles} depicts the velocity profiles associated with both the optimality conditions and the perturbed ones, computed via a numeric finite-difference scheme.
Specifically, referring to the optimal parameter values for the SRB model,
the flow solution is characterized by a cross-sectional average velocity of
$\bar{V} \approx 20$ $\text{mm/s}$, corresponding to a flow shear-rate measure
$\dot{\gamma}_F=2 \bar{V} / R \approx 200$
$\text{s}^{-1}$. The Carreau numbers \cite{tabakova2020oscillatory, shahsavari2015mobility},
computed with respect to the characteristic shear rates $\dot{\gamma}_0=1/\lambda_0$
and $\dot{\gamma}_\infty=1/\lambda_\infty$, are
$\text{Cu}_0 = \dot{\gamma}_F/\dot{\gamma}_0 = 19.17$ and
$\text{Cu}_\infty = \dot{\gamma}_F/\dot{\gamma}_\infty = 0.18$,
indicating an intermediate shear-thinning regime.
Moreover, Fig. \ref{fig:vel profiles} also shows the analytical solutions derived by assuming the fluid response in the cylindrical nozzle as completely described by the Ostwald-de Waele power-law model, that is by considering $\mu(\dot{\gamma})=K\dot{\gamma}^{N-1}$. In this case, the velocity profile is analytically represented by \cite{bird1987dynamics,conti2022models}:
\begin{align}
     &v_z(r)=\frac{N}{N+1}\left(\frac{\Delta p}{2LK} \right)^{\frac{1}{N}}\left(R^\frac{N+1}{N}-r^\frac{N+1}{N}\right)\,,
    \label{eq:analyticalPL}
\end{align}
where, accounting for Eq. \eqref{eq:K_muratio_PWA}, model parameters $K$ and $N$ are computed via $\mu_0$, $\lambda^\ast$ and $N=n^\ast$ (with $n^\ast=\eta$ and $\lambda^\ast=\lambda_0$ for SRB; $n^\ast=n$ and $\lambda^\ast=\lambda$ for BCCY) corresponding to the results of the inverse identification procedure (see Table \ref{tab:comparison SRB and BCCY pow law flow}).

The proposed results show that:

\begin{itemize}
    \item The velocity profiles numerically computed using both the SRB and BCCY models under optimal conditions are very similar, with differences of less than 2\%. Moreover, both models yield nearly the same cross-section average velocity $\bar{V}$.

    \item When model parameters deviate from their optimal values, the BCCY formulation continues to provide a reasonably accurate rheological description compared to the reference data, unlike the SRB formulation (Fig. \ref{fig:synth_rheol}). Specifically, the maximum deviation between BCCY predictions and the reference data remains below 11\%, whereas discrepancies in the SRB-based rheological description can exceed 60\%.

    \item Although significant deviations from the optimal model parameters do not substantially affect the rheological response of the BCCY model, the resulting flow description can differ significantly from the actual one (see Fig. \ref{fig:vel profiles}). Specifically, considerable discrepancies arise when comparing the BCCY-based velocity profiles computed using the optimal and the perturbed parameters, with differences exceeding 17\% (approximately 14\% in the average velocity value). This evidence further confirms the identifiability issue of the BCCY model, as highlighted by Gallagher et al. \cite{gallagher2019non}.
    Conversely, since deviations from the optimal parameter values lead to
    significant changes in the rheological response described by the SRB model
    (see Fig. \ref{fig:synth_rheol}), the corresponding velocity profiles
    coherently exhibit significant discrepancies (of approximately $40\%$).

    \item The analytical velocity profile in Eq. \eqref{eq:analyticalPL} obtained using the Ostwald–de Waele power-law model based on the SRB nonlinear regression exhibits excellent agreement with the optimal numerical solutions, with differences lower than 4\%. In contrast, the power-law analytical profile derived by employing model parameters obtained from the BCCY nonlinear regression shows significant discrepancies compared to the optimal numerical solutions, with differences exceeding 17\%. Given the values summarized in Table \ref{tab:comparison SRB and BCCY pow law flow}, this finding further confirms that the BCCY model parameter $n$ cannot be directly adopted—or straightforwardly identified—within a power-law description. Conversely, the SRB model parameter $\eta$ is better suited to approximate the power-law index $N$. Therefore, the SRB formulation effectively overcomes the critical identifiability issues of the BCCY model, enabling a simple and efficient estimation procedure based on a direct approach.
\end{itemize}

\section{Conclusions} 
\label{sec:conclusions}

The Carreau-Yasuda model (BCCY) is widely used to describe the shear-thinning behaviour of non-Newtonian inelastic fluids in many research and industrial applications. However, its parameter identification through experimental data fitting is often affected by intrinsic identifiability issues, leading to misleading interpretations of model parameters and unreliable flow predictions. To address these limitations, this paper proposed a novel rheological formulation specifically designed to mitigate such identifiability challenges.

Comparisons based on analytical arguments and numerical assessments demonstrated that the proposed model relies on physically meaningful parameters, whose identifiability is not compromised by the key issues affecting the Carreau-Yasuda formulation. The new approach allows for effective parameter estimation through a straightforward direct identification strategy, eliminating the need for numerical inverse
optimization procedures based on nonlinear regression techniques.
Moreover, the proposed formulation enables the identication of two Carreau
numbers based on the two characteristic shear rates of the fluid.

The results presented in this study highlight that the novel model not only provides
a robust and accurate description of shear-thinning fluids but also ensures improved
reliability in flow predictions. By addressing the fundamental shortcomings of the
BCCY model, this formulation represents a promising alternative for advanced applications
requiring precise rheological characterizations.

Future work may include the extension of the analysis to more complex flow conditions,
the comparison with other generalized Newtonian fluid (GNF) formulations,
and the validation of the model against a broader range of experimental data.
In particular, the identifiability properties of the proposed rheological
relationship could be further investigated by performing uncertainty quantification
on both the rheological parameters and the flow model parameters
(e.g., geometry, reference velocity), as carried out by Kim et al. \cite{kim2019uncertainty}
for other GNF models.
Moreover, potential applications span various fields,
including industrial processes and biomedical systems, with particular relevance to bioprinting technologies \cite{chirianni2024development, chirianni2024influence, conti2022models}
and the modelling of cell-laden hydrogels \cite{gaziano2024computational, gaziano2024phase},
where accurate prediction of shear-thinning behaviour is critical to preserve cell
viability and achieve precise material deposition. The model also holds promise
for advanced hemodynamic applications, for instance related to the influence of
blood-vessel interaction in health and pathological tissue remodelling \cite{gierig2024post}.

\section*   {Acknowledgments}

Part of this work was carried out with the support from the Italian National Group for Mathematical Physics GNFM-INdAM.

Giuseppe Vairo acknowledges financial support by the Italian Ministry of University and Research (MUR) under the National Recovery and Resilience Plan (NRRP), within the PRIN 2022 program, Project 2022T3SLAZ, CUP E53D23003700006.
Michele Marino acknowledges financial support under the National Recovery and Resilience Plan (NRRP) by the Italian Ministry of University and Research (MUR), funded by the European Union - NextGenerationEU, within the PRIN 2022 program, Project 2022Z24WLR (project acronym: MATERIAL), CUP: E53C24002920006.
Franceco Viola acknowledges financial support by the European Research Council (ERC)
under the European Union’s Horizon Europe research and innovation program,
Project CARDIOTRIALS, Grant No. 101039657.

\appendix 
\section{}
\label{appendixA}

Consistently with symbols defined in Sections \ref{sec:BCCY_model} and \ref{sec:SRB_model},
and by assuming $\alpha=a$, $\eta=n=N$, $\lambda=\lambda_0$, let the following
notation be introduced
\begin{align}
    x=(\lambda_0 \dot{\gamma})^a,\qquad \beta=\frac{1-n}{a}, \qquad g=\left(\frac{\lambda_\infty}{\lambda_0}\right)^a,
\end{align}
so that
\begin{align}
 &\mu_\infty/\mu_0=g^\beta,\qquad 0<\beta<1,\qquad 0<g<g^\beta<1.
 \label{eq:limitazioni}
\end{align}
Thereby, the shear rate range $1/\lambda_0\le \dot{\gamma}\le 1/\lambda_\infty$, that
identifies the shear-thinning region, corresponds to $1<x<1/g$. Accordingly,
Eqs. \eqref{eq:ratioBCCY} and \eqref{eq:ratioSRB} can be recast as
\begin{align}
    &\mathcal{S}_\text{BCCY}(x)=\frac{x(1-g^\beta)}{(1+x)[(1-g^\beta)+g^\beta (1+x)^\beta]} \label{eq:SBCCYx}\ ,\\
    &\mathcal{S}_\text{SRB}(x)=\frac{x(1-g)}{(1+x)(1+gx)}\label{eq:SSRBx}\ .
\end{align}

\subsection{Proof of the inequality \eqref{eq:upperbound}}

For shear rate levels inducing shear-thinning (i.e., for $x>1$), functions
$(1+x)/x$ and $(1+x)^{\beta+1}/x$ are both strictly greater than one. Moreover,
the latter attains its minimum value $(1+\beta)^{1+\beta}/\beta^\beta>1$ at $x=1/\beta>1$.
As a result, and accounting for relationships \eqref{eq:limitazioni}, the following
inequality holds
\begin{align}
    0<\mathcal{S}_\text{BCCY}(x)&=\frac{1-g^\beta}{\frac{1+x}{x}\,(1-g^\beta)+ \frac{(1+x)^{\beta+1}}{x}g^\beta}\nonumber\\
    &<\frac{1-g^\beta}{1+g^\beta\left[ \frac{(1+\beta)^{\beta+1}}{\beta^\beta}-1\right] }<1 \ ,
\end{align}
that corresponds to the inequality \eqref{eq:upperbound}.

\subsection{Proof of the inequality \eqref{eq:DSg0}}

The derivative of Eq. \eqref{eq:SSRBx} with respect to $x$ gives
\begin{align}
    \frac{d\mathcal{S}_\text{SRB}}{dx}=\frac{(1-t)(1-x^2t)}{(1+x)^2(1+xt)^2}\ ,
\end{align}
resulting in $\text{sgn}(d \mathcal{S}_\text{SRB}/dx) = \text{sgn}(1-x^2t)$.
Since the function $(1-x^2t)$ is strictly decreasing for $x>1$ and it vanishes
for $x=1/\sqrt{t}$, the corresponding stationary condition for $\mathcal{S}_\text{SRB}$
corresponds to the maximum value
\begin{align}
&\mathcal{S}_\text{SRB}\left(\frac{1}{\sqrt{t}}\right)=(1-\sqrt{t})/(1+\sqrt{t}),
\end{align}
that is the Eq. \eqref{eq:SSRBmax}.

In order to prove the inequality \eqref{eq:DSg0}, let the function
$\Delta\mathcal{S}(x)=\mathcal{S}_\text{SRB}(x)-\mathcal{S}_\text{BCCY}(x)$ be
introduced. Since the strict positivity of both the denominators of $\mathcal{S}_\text{BCCY}$
and $\mathcal{S}_\text{SRB}$ when $1<x<1/g$, the sign of $\Delta\mathcal{S}$
coincides with the sign of the following continuous function
\begin{align}
    F(x)=(1-g)[(1-g^\beta)+g^\beta (1+x)^\beta]-(1-g^\beta)(1+gx)\ .
\end{align}
It can be simply shown that the  solutions of the equation $F(x)=0$ are
\begin{align}
    x_1=-1,\qquad x_2 =\frac{1}{g}\left(\frac{1-g}{1-g^\beta}\right)^{\frac{1}{1-\beta}}-1\ .
\end{align}

Moreover, due to inequalities in \eqref{eq:limitazioni}, the following condition holds
\begin{align}
    \frac{1-g}{1-g^\beta}>\frac{1-g^2}{1-g^\beta}=\frac{1-g}{1-g^\beta}\,(1+g)>1+g>(1+g)^{1-\beta}\ ,
\end{align}
that straight implies $x_2  >1/g>1$. 

On the other hand, due to the continuity of $F(x)$ and observing that $F(0)=g^\beta -g>0$,
it results
\begin{align}
    \text{sgn}(F)=\text{sgn}(\Delta \mathcal{S})=+1 \qquad \text{for } 1<x<\frac{1}{g}\ ,
\end{align}
that is equivalent to the inequality \eqref{eq:DSg0}.

% \nocite{*}
% \section{}
\bibliography{mybibfile2}

\begin{thebibliography}{10}
\expandafter\ifx\csname url\endcsname\relax
  \def\url#1{\texttt{#1}}\fi
\expandafter\ifx\csname urlprefix\endcsname\relax\def\urlprefix{URL }\fi
\expandafter\ifx\csname href\endcsname\relax
  \def\href#1#2{#2} \def\path#1{#1}\fi

\bibitem{johnston2004non}
B.~M. Johnston, P.~R. Johnston, S.~Corney, D.~Kilpatrick, Non-newtonian blood
  flow in human right coronary arteries: steady state simulations, Journal of
  biomechanics 37~(5) (2004) 709--720.

\bibitem{amorim2021insights}
P.~A. Amorim, M.~d’{\'A}vila, R.~Anand, P.~Moldenaers, P.~Van~Puyvelde,
  V.~Bloemen, Insights on shear rheology of inks for extrusion-based 3d
  bioprinting, Bioprinting 22 (2021) e00129.

\bibitem{sauty2022enabling}
B.~Sauty, G.~Santesarti, T.~Fleischhammer, P.~Lindner, A.~Lavrentieva,
  I.~Pepelanova, M.~Marino, Enabling technologies for obtaining desired
  stiffness gradients in gelma hydrogels constructs, Macromolecular Chemistry
  and Physics 223~(2) (2022) 2100326.

\bibitem{chhabra2011non}
R.~P. Chhabra, J.~F. Richardson, Non-Newtonian flow and applied rheology:
  engineering applications, Butterworth-Heinemann, 2011.

\bibitem{bird1987dynamics}
R.~B. Bird, R.~C. Armstrong, O.~Hassager, Dynamics of polymeric liquids. Vol.
  1: Fluid mechanics, John Wiley and Sons Inc., New York, NY, 1987.

\bibitem{rao2010rheology}
M.~A. Rao, Rheology of fluid and semisolid foods: principles and applications,
  Springer Science \& Business Media, 2010.

\bibitem{cherry2013shear}
E.~M. Cherry, J.~K. Eaton, Shear thinning effects on blood flow in straight and
  curved tubes, Physics of Fluids 25~(7) (2013).

\bibitem{gijsen1999influence}
F.~J. Gijsen, F.~N. van~de Vosse, J.~Janssen, The influence of the
  non-newtonian properties of blood on the flow in large arteries: steady flow
  in a carotid bifurcation model, Journal of biomechanics 32~(6) (1999)
  601--608.

\bibitem{poole2023inelastic}
R.~J. Poole, Inelastic and flow-type parameter models for non-newtonian fluids,
  Journal of Non-Newtonian Fluid Mechanics 320 (2023) 105106.

\bibitem{waele1923viscometry}
A.~Waele, Viscometry and plastometry, Oil and Colour Chemists' Association,
  1923.

\bibitem{ostwald1925ueber}
W.~Ostwald, Ueber die geschwindigkeitsfunktion der viskosit{\"a}t disperser
  systeme. i, Kolloid-Zeitschrift 36~(2) (1925) 99--117.

\bibitem{carreau1972rheological}
P.~J. Carreau, Rheological equations from molecular network theories,
  Transactions of the Society of Rheology 16~(1) (1972) 99--127.

\bibitem{matsuhisa1965analytical}
S.~Matsuhisa, R.~B. Bird, Analytical and numerical solutions for laminar flow
  of the non-newtonian ellis fluid, AIChE Journal 11~(4) (1965) 588--595.

\bibitem{cross1965rheology}
M.~M. Cross, Rheology of non-newtonian fluids: a new flow equation for
  pseudoplastic systems, Journal of colloid science 20~(5) (1965) 417--437.

\bibitem{yasuda1979investigation}
K.~Yasuda, Investigation of the analogies between viscometric and linear
  viscoelastic properties of polystyrene fluids, Ph.D. thesis, Massachusetts
  Institute of Technology (1979).

\bibitem{gallagher2019non}
M.~T. Gallagher, R.~A. Wain, S.~Dari, J.~P. Whitty, D.~J. Smith,
  Non-identifiability of parameters for a class of shear-thinning rheological
  models, with implications for haematological fluid dynamics, Journal of
  Biomechanics 85 (2019) 230--238.

\bibitem{mazzanti2016rheological}
V.~Mazzanti, F.~Mollica, N.~El~Kissi, Rheological and mechanical
  characterization of polypropylene-based wood plastic composites, Polymer
  Composites 37~(12) (2016) 3460--3473.

\bibitem{bair2006more}
S.~Bair, A more complete description of the shear rheology of high-temperature,
  high-shear journal bearing lubrication, Tribology transactions 49~(1) (2006)
  39--45.

\bibitem{meza2021effect}
B.~E. Meza, J.~M. Peralta, S.~E. Zorrilla, Effect of temperature and
  composition on rheological behaviour and sagging capacity of glaze materials
  for foods, Food Hydrocolloids 117 (2021) 106689.

\bibitem{godfrey1985identifiability}
K.~Godfrey, J.~DiStefano~III, Identifiability of model parameter, IFAC
  Proceedings Volumes 18~(5) (1985) 89--114.

\bibitem{bellman1970structural}
R.~Bellman, K.~J. {\AA}str{\"o}m, On structural identifiability, Mathematical
  biosciences 7~(3-4) (1970) 329--339.

\bibitem{guillaume2019introductory}
J.~H. Guillaume, J.~D. Jakeman, S.~Marsili-Libelli, M.~Asher, P.~Brunner,
  B.~Croke, M.~C. Hill, A.~J. Jakeman, K.~J. Keesman, S.~Razavi, et~al.,
  Introductory overview of identifiability analysis: A guide to evaluating
  whether you have the right type of data for your modeling purpose,
  Environmental Modelling \& Software 119 (2019) 418--432.

\bibitem{raue2009structural}
A.~Raue, C.~Kreutz, T.~Maiwald, J.~Bachmann, M.~Schilling, U.~Klingm{\"u}ller,
  J.~Timmer, Structural and practical identifiability analysis of partially
  observed dynamical models by exploiting the profile likelihood,
  Bioinformatics 25~(15) (2009) 1923--1929.

\bibitem{santesarti2024quasi}
G.~Santesarti, M.~Marino, F.~Viola, R.~Verzicco, G.~Vairo, A quasi-analytical
  solution for “carreau-yasuda-like” shear-thinning fluids flowing in
  slightly tapered pipes, Journal of Non-Newtonian Fluid Mechanics, (submitted)
  \href{https://arxiv.org/abs/2502.14991}{\textcolor{blue}{arXiv:2502.14991}}
  (2025).

\bibitem{tabakova2020oscillatory}
S.~Tabakova, N.~Kutev, S.~Radev, Oscillatory carreau flows in straight
  channels, Royal Society Open Science 7~(5) (2020) 191305.

\bibitem{shahsavari2015mobility}
S.~Shahsavari, G.~H. McKinley, Mobility of power-law and carreau fluids through
  fibrous media, Physical Review E 92~(6) (2015) 063012.

\bibitem{itskov2007tensor}
M.~Itskov, et~al., Tensor algebra and tensor analysis for engineers, Springer,
  2007.

\bibitem{irgens2014rheology}
F.~Irgens, Rheology and non-newtonian fluids, Vol. 190, Springer, 2014.

\bibitem{zare2019analysis}
Y.~Zare, S.~P. Park, K.~Y. Rhee, Analysis of complex viscosity and shear
  thinning behavior in poly (lactic acid)/poly (ethylene oxide)/carbon
  nanotubes biosensor based on carreau--yasuda model, Results in Physics 13
  (2019) 102245.

\bibitem{ohta2005dynamic}
M.~Ohta, E.~Iwasaki, E.~Obata, Y.~Yoshida, Dynamic processes in a deformed drop
  rising through shear-thinning fluids, Journal of non-newtonian fluid
  mechanics 132~(1-3) (2005) 100--107.

\bibitem{pratumwal2017whole}
Y.~Pratumwal, W.~Limtrakarn, S.~Muengtaweepongsa, P.~Phakdeesan,
  S.~Duangburong, P.~Eiamaram, K.~Intharakham, Whole blood viscosity modeling
  using power law, casson, and carreau yasuda models integrated with image
  scanning u-tube viscometer technique., Songklanakarin Journal of Science \&
  Technology 39~(5) (2017).

\bibitem{colosi2016microfluidic}
C.~Colosi, S.~R. Shin, V.~Manoharan, S.~Massa, M.~Costantini, A.~Barbetta,
  M.~R. Dokmeci, M.~Dentini, A.~Khademhosseini, Microfluidic bioprinting of
  heterogeneous 3d tissue constructs using low-viscosity bioink, Advanced
  materials 28~(4) (2016) 677--684.

\bibitem{aho2013foundational}
K.~A. Aho, Foundational and applied statistics for biologists using R, CRC
  Press, 2013.

\bibitem{motulsky2004fitting}
H.~Motulsky, A.~Christopoulos, Fitting models to biological data using linear
  and nonlinear regression: a practical guide to curve fitting, Oxford
  University Press, 2004.

\bibitem{singh2019fitting}
P.~K. Singh, J.~M. Soulages, R.~H. Ewoldt, On fitting data for parameter
  estimates: residual weighting and data representation, Rheologica Acta 58
  (2019) 341--359.

\bibitem{bailer1990note}
A.~J. Bailer, C.~J. Portier, A note on fitting one-compartment models:
  Non-linear least squares versus linear least squares using transformed data,
  Journal of Applied Toxicology 10~(4) (1990) 303--306.

\bibitem{freund2015quantitative}
J.~B. Freund, R.~H. Ewoldt, Quantitative rheological model selection: Good fits
  versus credible models using bayesian inference, Journal of Rheology 59~(3)
  (2015) 667--701.

\bibitem{cornell1987factors}
J.~Cornell, R.~Berger, Factors that influence the value of the coefficient of
  determination in simple linear and nonlinear regression models,
  Phytopathology 77~(1) (1987) 63--70.

\bibitem{cameron1997r}
A.~C. Cameron, F.~A. Windmeijer, An r-squared measure of goodness of fit for
  some common nonlinear regression models, Journal of econometrics 77~(2)
  (1997) 329--342.

\bibitem{taylor1997introduction}
J.~Taylor, Introduction to error analysis, the study of uncertainties in
  physical measurements, 1997.

\bibitem{jones2019beyond}
A.~G. Jones, Beyond chi-squared: additional measures of the closeness of a
  model to data, ASEG Extended Abstracts 2019~(1) (2019) 1--6.

\bibitem{laborda2021feature}
J.~Laborda, S.~Ryoo, Feature selection in a credit scoring model, Mathematics
  9~(7) (2021) 746.

\bibitem{shaw2017uncertainty}
B.~D. Shaw, Uncertainty analysis of experimental data with R, Chapman and
  Hall/CRC, 2017.

\bibitem{chirianni2024influence}
F.~Chirianni, G.~Vairo, M.~Marino, Influence of extruder geometry and bio-ink
  type in extrusion-based bioprinting via an in silico design tool, Meccanica
  59~(8) (2024) 1285--1299.

\bibitem{chirianni2024development}
F.~Chirianni, G.~Vairo, M.~Marino, Development of process design tools for
  extrusion-based bioprinting: From numerical simulations to nomograms through
  reduced-order modeling, Computer Methods in Applied Mechanics and Engineering
  419 (2024) 116685.

\bibitem{conti2022models}
M.~Conti, G.~Santesarti, F.~Scocozza, M.~Marino, Models and simulations as
  enabling technologies for bioprinting process design, in: Bioprinting,
  Elsevier, 2022, pp. 137--206.

\bibitem{kim2019uncertainty}
J.~Kim, P.~K. Singh, J.~B. Freund, R.~H. Ewoldt, Uncertainty propagation in
  simulation predictions of generalized newtonian fluid flows, Journal of
  Non-Newtonian Fluid Mechanics 271 (2019) 104138.

\bibitem{gaziano2024computational}
P.~Gaziano, M.~Marino, Computational modeling of cell motility and clusters
  formation in enzyme-sensitive hydrogels, Meccanica 59~(8) (2024) 1335--1349.

\bibitem{gaziano2024phase}
P.~Gaziano, M.~Marino, A phase-field model of cell motility in biodegradable
  hydrogel scaffolds for tissue engineering applications, Computational
  Mechanics 74~(1) (2024) 45--66.

\bibitem{gierig2024post}
M.~Gierig, P.~Gaziano, P.~Wriggers, M.~Marino, Post-angioplasty remodeling of
  coronary arteries investigated via a chemo-mechano-biological in silico
  model, Journal of Biomechanics 166 (2024) 112058.

\end{thebibliography}

\end{document}